\documentclass[onecolumn,secnumarabic,amssymb,amsmath, nobibnotes, aps]{revtex4-2}
\usepackage{graphicx}

\begin{document}

\title{Quantum Simulation of \\
  Energy Bifurcation and $\mathbb{Z}_{2}$-Symmetry Restoration in Macroscopic Quantum Tunneling}%

\author{Masao Hirokawa}%
\email[]{hirokawa@inf.kyushu-u.ac.jp}
\altaffiliation{Graduate School of Information Science and Electrical Engineering, Kyushu University, 744 Motooka, Nishi-ku, Fukuoka, 819-0395, Japan}
\altaffiliation{Quantum Computing System Center, Kyushu University, 744 Motooka, Nishi-ku, Fukuoka, 819-0395, Japan}
\altaffiliation{Quantum and Spacetime Research Institute, Kyushu University, 744 Motooka, Nishi-ku, Fukuoka, 819-0395, Japan}
\date{\today}%

\begin{abstract}
  Macroscopic quantum tunneling (MQT),
  a cornerstone of Leggett's program,
  is deeply linked with instanton physics,
  yet its experimental verification remains elusive.
  This Perspective demonstrates that the quantum Rabi model
  manifests observable, instanton-like effects via quantum simulation.
  In the MQT regime, qubit-boson interactions drive Polyakov's energy bifurcation,
  governing tunneling and spontaneous symmetry breaking.
  Mapping the quantum Rabi model onto an effective double-well potential
  reveals that while tunneling suppression induces spontaneous symmetry breaking,
  instanton-like contributions act to restore it.
  This mechanism enables experimental access to the classical Euclidean action
  of an effective instanton-like particle,
  offering a route to probe non-perturbative phenomena.
\end{abstract}


\maketitle

\section{Introduction}
\label{sec:intro}

The investigation of macroscopic quantum phenomena
is often referred to as Leggett's program \cite{leg80,leg87,takagi02},
with macroscopic quantum tunneling (MQT) \cite{cl81,dev84,mar85,dev85}
serving as a prime example. 
The quantum Rabi model can be regarded as a representation of MQT
via a unitary transformation,
since it models the two-level approximation of tunneling 
between two symmetric wells in a double-well potential
(see Section \ref{sec:qrm_2lsa}).
In light of the MQT features revealed by this model,
it is instructive to revisit Polyakov's assertion in Ref.\cite{pol77} (p.430)
regarding how instantons emerge to restore spontaneously broken symmetry.
This emergence is observed in the full process of energy splitting
of the degenerate ground state.
Indeed, an instanton-like effect emerges in the ground state energy of the quantum Rabi model \cite{hir25},
which challenges his assertion (see Section 4).
This model can often be realized in quantum simulators \cite{braumueller17,cai21,cai22,wu24}.
In this context, quantum simulation refers to the implementation of
a quantum phenomenon with a programmable quantum system.
The technology of quantum simulators has been advancing
continuously over the last 20 years
\cite{ger11,end12,yan16,est16,braumueller17,kok19,sch19,yan20,cai21,cai22,zha22,wu24,wehinger25,cuadra25}.

Consider the ground state of a macroscopic quantum system with
a symmetric double-well potential 
separated by a barrier.
We assume that the Hamiltonian of the macroscopic system possesses symmetry,
hereafter referred to as the \textit{macrosystem Hamiltonian} in accordance with Ref.\cite{takagi02}. 
The penetration (and thus tunneling) is absent as the barrier between the two wells becomes infinitely high.
As is well known, the ground state becomes degenerate, implying that the symmetry is spontaneously broken.
According to Polyakov, instantons emerge to restore the symmetry by enabling tunneling between the wells.
This restoration lifts the degeneracy, splitting the ground states into
a new ground state and a first excited state. 
In this paper, 
we refer to this energy splitting as \textit{Polyakov's energy bifurcation}.

\begin{figure}[h]
\centering
\includegraphics[width=0.30\textwidth]{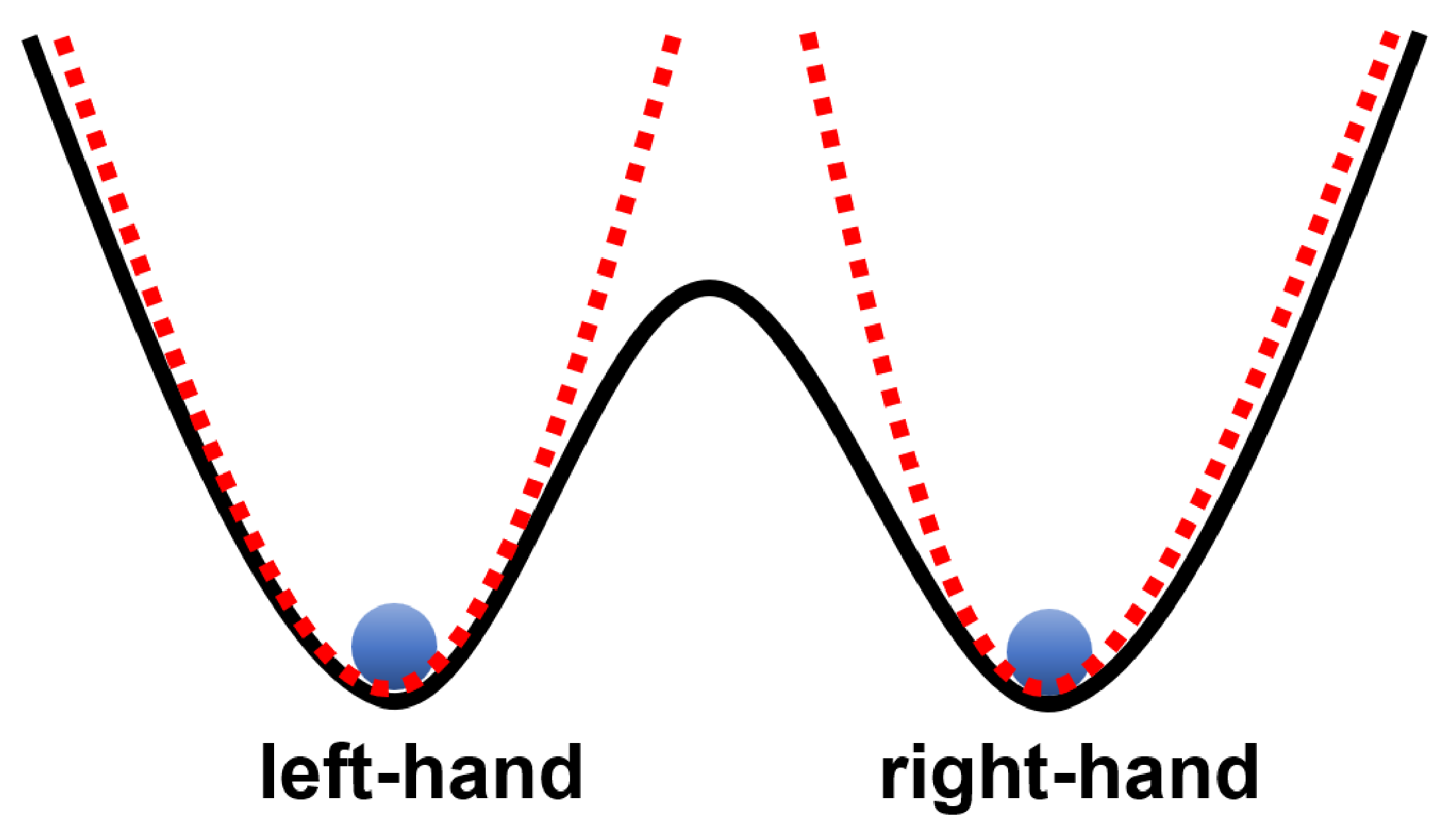}
\caption{The graph shows a double-well potential $V(x)$.} 
\label{fig:double-well_potential_1}
\end{figure}

The macrosystem Hamiltonian is described by the Schr\"{o}dinger operator
with the symmetric double-well potential $V(x)$ \cite{takagi02}:  
\begin{align}
H_{\mbox{\tiny{MS}}}
=\frac{1}{2}\hat{p}^{2}+V(\hat{x}),
  \label{eq:ms-Hamiltonian}
\end{align}
where $\hat{x}$ and $\hat{p}$
are respectively the operators of position $x$ and momentum $p$
satisfying the canonical commutation relation,
$[\hat{x} , \hat{p}]=i\hbar$.
The potential $V(x)$ has two symmetrical wells.
We assume that the minima of the wells are separated, 
as shown in Fig.\ref{fig:double-well_potential_1},
and that the quantum motion of the particle in each well is semiclassical. 
While this physical system is called an ``intrinsically two-state system'' in Ref.\cite{leg87}, 
we call it a \textit{two-level system} in this paper. 
The states of the semiclassical particle in the left-hand
and right-hand wells are respectively approximated by two states of the two-level system:
the ground state and the first excited state.
Systems with two macroscopic states, like ammonia molecules,
are often described using a two-level system
(see III-10-1 of Ref.\cite{feynman3}, as well as Refs.\cite{hund27a, hund27b, hund27c, wightman89, leg87, th09, takagi02}).  
Thus, we consider this two-level system to represent a particle localized
in either of the two individual minima.
In Ref.\cite{leg87}, by choosing the basis such that the eigenstates of the  $Z$-gate $\sigma_{z}$
correspond to the states localized in the left and right wells,
the energy of the two-level system is described by the Hamiltonian
$-\, \frac{\hbar\omega_{\mathrm{a}}}{2}\sigma_{x}+\frac{1}{2}\varepsilon\sigma_{z}$. 
The first term describes tunneling between the two wells mediated by
the $X$-gate $\sigma_{x}$, whereas the second term introduces an energy bias.

\begin{figure}[h]
\centering
\includegraphics[width=0.30\textwidth]{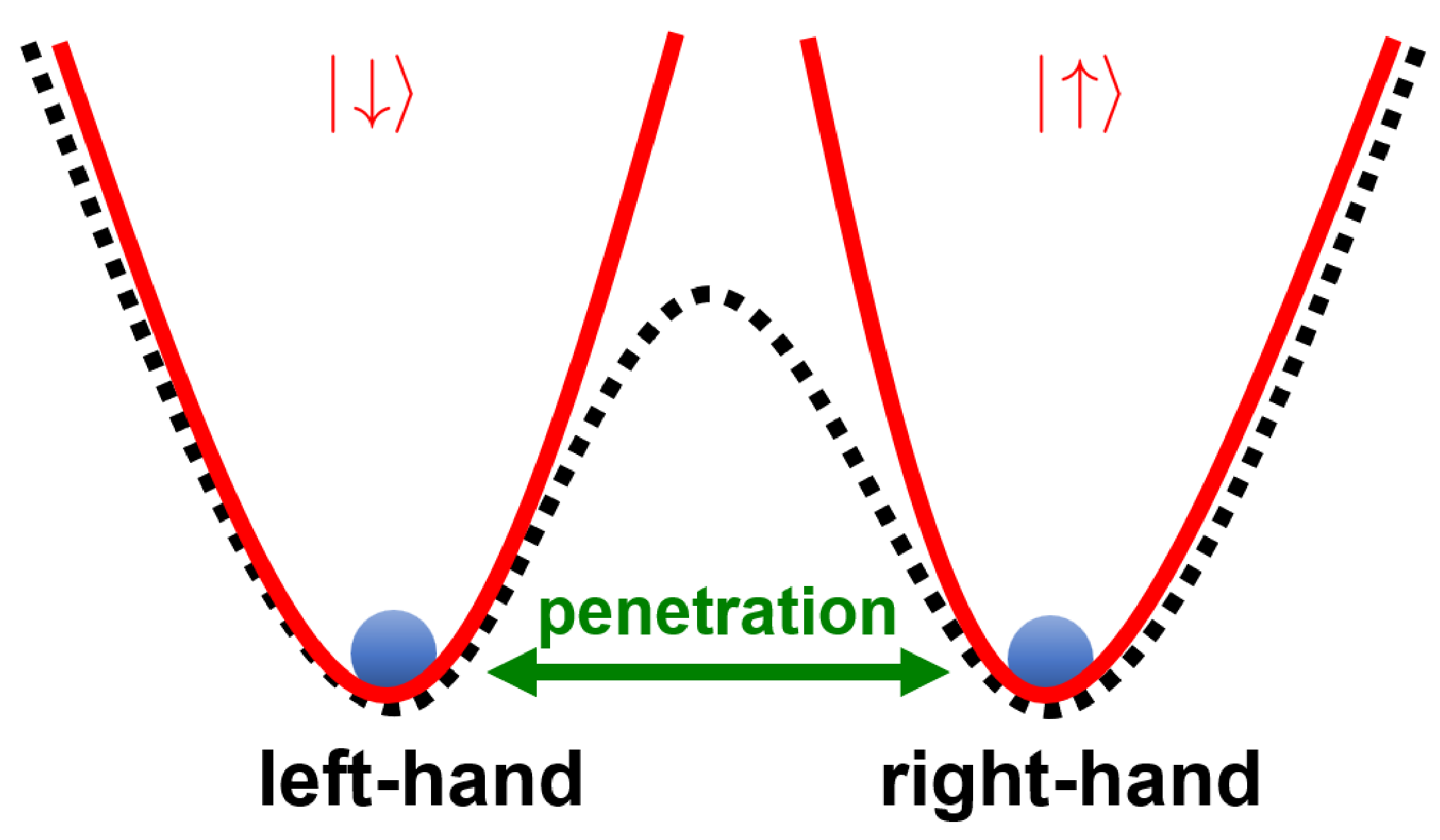}
\caption{This figure illustrates a two-level approximation of the double-well potential $V(x)$.
  The Hamiltonian for the red part is given by
  $\begin{pmatrix}
    0 & 0 \\
    0 & 1
  \end{pmatrix}
  \otimes H_{\mbox{\tiny{MS}}}^{\mathrm{ho}}(\omega_{\mathrm{c}}, -x_{0})$
  and
  $\begin{pmatrix}
    1 & 0 \\
    0 & 0
  \end{pmatrix}
  \otimes H_{\mbox{\tiny{MS}}}^{\mathrm{ho}}(\omega_{\mathrm{c}}, +x_{0})$.
  The penetration (shown in green) is formed as described in Section \ref{sec:qrm_2lsa}.} 
\label{fig:double-well_potential_2}
\end{figure}

Let us consider a double-well potential $V(x)$
with two symmetrical wells having minima
at $x=\pm x_{0}$ (see Fig.\ref{fig:double-well_potential_1}).  
Following Ref.\cite{takagi02}, we approximate
the potential $V(x)$ as
$V_{x_{0}}(x)=
\frac{\omega_{\mathrm{c}}^{2}}{2}(x\mp x_{0})^{2}$
near $x=\pm x_{0}$,
where $\omega_{\mathrm{c}}$ is the frequency of the single-mode boson.
Thus, for $x$ sufficiently close to $\pm x_{0}$,
the macrosystem Hamiltonian $H_{\mbox{\tiny{MS}}}$ can be approximated as
\begin{align*}
H_{\mbox{\tiny{MS}}}
\approx \frac{1}{2}\hat{p}^{2}+V_{x_{0}}(\hat{x}).
  \label{eq:approx_H_MS}
\end{align*}
We denote the right-hand side of this approximation
by $H_{\mbox{\tiny{MS}}}^{\mathrm{ho}}(\omega_{\mathrm{c}}, \pm x_{0})$. 
However, approximating the original macrosystem Hamiltonian $H_{\mbox{\tiny{MS}}}$
solely with the harmonic oscillator Hamiltonians
$H_{\mbox{\tiny{MS}}}^{\mathrm{ho}}(\omega_{\mathrm{c}}, \pm x_{0})$
is insufficient,
since the harmonic oscillator potentials confine the particles.
Thus, an approach is required
to account for the tunneling between the two wells. 
Meanwhile, the quantum Rabi model, as well as the spin-boson model,
has been shown to exhibit properties similar to tunneling particles
in a double-well potential
\cite{hir99,irish14,hir25}.
Therefore, we develop the approach by mapping the quantum Rabi model onto an effective double-well potential,
rendering the well-known analogy with the ground-state energy bifurcation explicit
(see Fig.\ref{fig:double-well_potential_2} and Section \ref{sec:qrm_2lsa}).
While tracking the full process of ground-state energy bifurcation for 
a Schr\"{o}dinger operator with a double-well potential \cite{col77,CC77,col85}
is purely theoretical,
quantum simulation using the quantum Rabi model could enable its experimental observation.

\begin{figure}[h]
\centering
\includegraphics[width=0.45\textwidth]{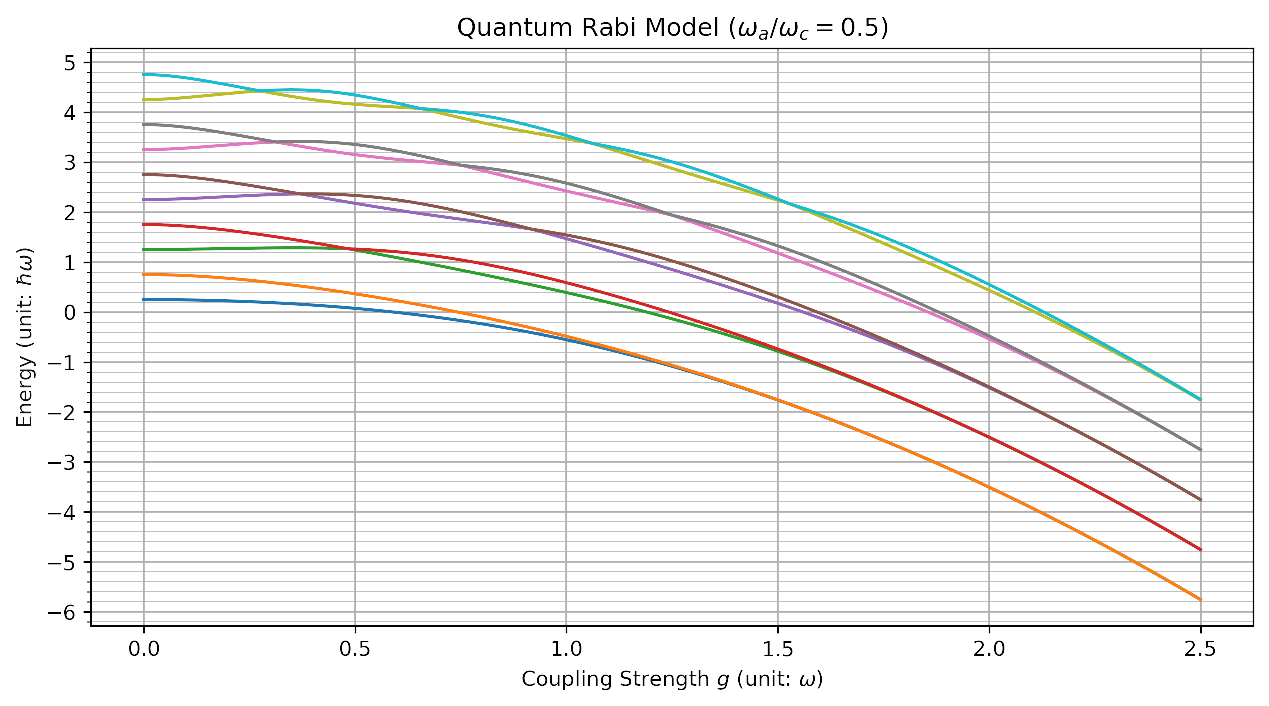}\qquad 
\includegraphics[width=0.45\textwidth]{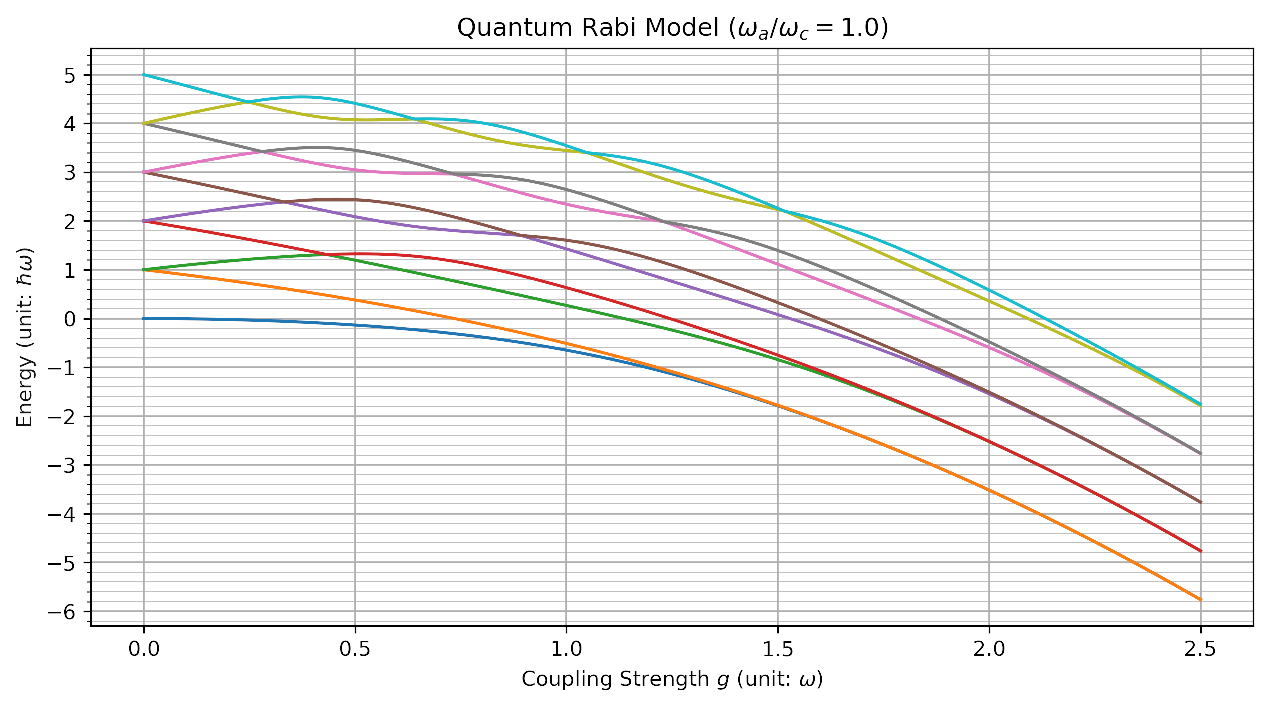}
\caption{Energy levels of the lowest ten states in the quantum Rabi model as a function of the coupling strength $\mathrm{g}$.
  The left and right panels show the cases for $\omega_{\mathrm{a}}=0.5\omega$, $\omega_{\mathrm{c}}=\omega$
  and $\omega_{\mathrm{a}}=\omega_{\mathrm{c}}=\omega$, respectively,
  where $\omega$ is a constant frequency.
  Here, $\omega_{\mathrm{a}}$ denotes the qubit transition frequency and
  $\omega_{\mathrm{c}}$ is the frequency of the single-mode boson.} 
\label{fig:qr_energy_0}
\end{figure}

The quantum Rabi model exhibits stable spontaneous $\mathbb{Z}_{2}$-symmetry breaking
in a certain limit (see Section \ref{sec:qrm_ssb}).
In the specific case where the qubit transition and the single-mode boson frequencies are strictly matched,
the $\mathbb{Z}_{2}$-symmetry breaking transition corresponds to the transition from $\mathcal{N}=2$ supersymmetry (SUSY)
to its spontaneous breaking \cite{hir15, hir24}.
As shown in Fig.\ref{fig:qr_energy_0}, in the strong-coupling limit
($\mathrm{g}\to\infty$), where $\mathrm{g}$ is the coupling strength between the qubit and the single-mode boson,
all eigenstates become nearly two-fold degenerate.
Furthermore, the eigenenergies are spaced at almost equal intervals
and diverge toward negative infinity, following a common parabolic
trend of $-c_{0}\mathrm{g}^{2}$.
The constant $c_{0}$ is determined by the self-energy of the single-mode boson's self-interaction
(see Section \ref{sec:qrm_ssb}).
This behavior results in spontaneous $\mathbb{Z}_{2}$-symmetry breaking.
The right panels of Fig.\ref{fig:qr_energy_0} show
the special case where the qubit transition frequency and the single-mode boson frequency are equal.
At $\mathrm{g}=0$, the ground state energy is simple (i.e., the ground state is unique),
while all excited states are two-fold degenerate.
The eigenenergies are distributed at equal intervals,
resulting in $\mathcal{N}=2$ SUSY.
Viewed from the strong-coupling limit ($\mathrm{g}\to\infty$) where
spontaneous SUSY breaking occurs,
the $\mathcal{N}=2$ SUSY emerges at much higher energy levels,
as illustrated in Fig.\ref{fig:qr_energy_0}.
This transition has been mathematically proven for the renormalized quantum Rabi Hamiltonian \cite{hir15}
(see Section \ref{sec:qrm_ssb})
and experimentally demonstrated using an ion-trap simulator \cite{cai22}.

The contents of this paper 
are organized as follows:
In Section \ref{sec:qrm_2lsa},
we introduce a unitary transformation that maps the quantum Rabi model to
a unitarily equivalent representation, termed the \textit{transformed quantum Rabi model}. 
This mapping provides a physical interpretation of the unbiased ($\varepsilon=0$)
two-level system Hamiltonian with a qubit-boson interaction,
thereby positioning the quantum Rabi model as a two-level approximation of the double-well potential.
Building on the framework established in Refs.\cite{hund27a, hund27b, hund27c, cl81, cl83, leg87},
Ref.\cite{hir25} briefly introduces a two-level-system approximation
using the transformed quantum Rabi model. 
In Section \ref{sec:qrm_ssb}, we clarify the physical mechanism of the transition to
spontaneous $\mathbb{Z}_{2}$-symmetry breaking
in terms of the interaction in the transformed quantum Rabi model.
In Section \ref{sec:polyakov}, based on the results for the classical Euclidean action of
an effective instanton-like particle \cite{hir25},
we show that the transformed quantum Rabi model reveals Polyakov's energy bifurcation.
Furthermore,
by leveraging this bifurcation,
we propose a formula for experimentally observing the classical Euclidean action
of the effective instanton-like particle.

\section{Transformed Quantum Rabi Model and Effective Double-Well Potential}
\label{sec:qrm_2lsa}

The state space of a qubit is $\mathbb{C}^{2}$, the $2$-dimensional unitary space,
where $\mathbb{C}$ is the set of all the complex numbers.
The up-spin state and the down-spin state are denoted by
$\mid\uparrow\rangle=\bigl(
\begin{smallmatrix}
   1 \\
   0
\end{smallmatrix}
\bigl)$ 
and
$\mid\downarrow\rangle=\bigl(
\begin{smallmatrix}
   0 \\
   1
\end{smallmatrix}
\bigl)$, respectively.
The state space of a single-mode boson is given by the boson Fock space $\mathcal{F}_{\mathrm{b}}$,
spanned by all the Fock states $\mid\!\!n\rangle$, $n=0, 1, 2, \cdots$. 
We define the total state space of the system as the Hilbert space
$\mathcal{F}=\mathbb{C}^{2}\otimes\mathcal{F}_{\mathrm{b}}$.
The standard Pauli matrices are denoted by
$\sigma_{x}=\big(
\begin{smallmatrix}
   0 & 1 \\
   1 & 0
\end{smallmatrix}
\big)$,
$\sigma_{y}=\big(
\begin{smallmatrix}
   0 & -i \\
   i & 0
\end{smallmatrix}
\big)$,
and 
$\sigma_{z}=\big(
\begin{smallmatrix}
   1 & 0 \\
   0 & -1
\end{smallmatrix}
\big)$.
The annihilation and creation operators
of the single-mode boson are denoted by $a$ and $a^{\dagger}$, respectively.
We use the notation `$1$' for the $2\times 2$ identity matrix, 
the identity operator on $\mathcal{F}_{\mathrm{b}}$,
and the numerical value $1$.
We often omit the tensor product symbol `$\otimes$' and write
the state $\mid\!\!\sharp\rangle\otimes\!\mid\!\!\psi\rangle$
as $\mid\!\!\sharp\rangle\!\mid\!\!\psi\rangle$
for the states $\mid\!\!\sharp\rangle\in\mathbb{C}^{2}$ and $\mid\!\!\psi\rangle\in\mathcal{F}_{\mathrm{b}}$. 
Correspondingly, an operator $A\otimes B$ acting on $\mathcal{F}$ is abbreviated as $AB$,
where the operators $A$ and $B$ act on $\mathbb{C}^{2}$ and $\mathcal{F}_{\mathrm{b}}$, respectively. 
In addition, we often omit `$1\otimes$' and `$\otimes 1$,'
writing $A\otimes 1$ as $A$ and $1\otimes B$ as $B$.

The unitary equivalence, $\mathbb{C}^{2}\otimes\mathcal{F}_{\mathrm{b}}
\cong \mathcal{F}_{\mathrm{b}}\oplus\mathcal{F}_{\mathrm{b}}$,
obtained by the correspondence, 
$$
\mathbb{C}^{2}\otimes\mathcal{F}_{\mathrm{b}}\ni
{\mid\uparrow\rangle}\!\otimes{\!\mid\!\!\psi\rangle}+\!\mid\downarrow\rangle{\otimes\!\!\mid\!\!\varphi\rangle}
\longleftrightarrow
\begin{pmatrix}
   \mid\!\!\psi\rangle \\
   \mid\!\!\varphi\rangle
\end{pmatrix}
\in\mathcal{F}_{\mathrm{b}}\oplus\mathcal{F}_{\mathrm{b}},
$$
allows us to identify the operator,
$(\mid\uparrow\rangle\langle\uparrow\mid)\otimes B_{1}$
$+$
$(\mid\uparrow\rangle\langle\downarrow\mid)\otimes B_{2}$
$+$
$(\mid\downarrow\rangle\langle\uparrow\mid)\otimes B_{3}$
$+$
$(\mid\downarrow\rangle\langle\downarrow\mid)\otimes B_{4}$ 
acting on $\mathcal{F}$
with the operator matrix  
$\bigl(
\begin{smallmatrix}
  B_{1} & B_{2} \\
  B_{3} & B_{4}
\end{smallmatrix}
\big)
$
acting on $\mathcal{F}_{\mathrm{b}}\oplus\mathcal{F}_{\mathrm{b}}$,
where $B_{i}$, $i=1, 2, 3, 4$, are operators on $\mathcal{F}_{\mathrm{b}}$.

The Hamiltonian of the quantum Rabi model reads
\begin{align*}
H_{\mbox{\tiny{QR}}}
    =\frac{\hbar\omega_{\mathrm{a}}}{2}\sigma_{z}
    +\hbar\omega_{\mathrm{c}}\left(a^{\dagger}a+\frac{1}{2}\right)
    +\hbar\mathrm{g}\sigma_{x}\left(a+a^{\dagger}\right),
\end{align*}
characterized by the qubit transition frequency $\omega_{\mathrm{a}}$, 
the single-mode-boson frequency $\omega_{\mathrm{c}}$,
and the coupling strength $\mathrm{g}\ge 0$.
Here, $\hbar$ is the Planck constant divided by $2\pi$. 
Let us define the parity operator $P$, associated with the spin localization and the boson number,
by $P=\sigma_{z}(-1)^{a^{\dagger}a}$.
Then, the parity symmetry of the quantum Rabi Hamiltonian is immediately evident:  
\begin{equation}
  \left[H_{\mbox{\tiny{QR}}},P\right]=0.
\label{eq:parity-symmetry}
\end{equation}

By introducing the position operator (i.e., single-mode scalar field) 
$\hat{x}=\sqrt{\frac{\hbar}{2\omega_{\mathrm{c}}}}(a+a^{\dagger})$
and the momentum operator (i.e., the conjugate field)
$\hat{p}=-i\sqrt{\frac{\hbar\omega_{\mathrm{c}}}{2}}(a-a^{\dagger})$,
which act on the boson Fock space $\mathcal{F}_{\mathrm{b}}$,
the quantum Rabi Hamiltonian can be rewritten as
\begin{equation}
  H_{\mbox{\tiny{QR}}}
  =
  \frac{\hbar\omega_{\mathrm{a}}}{2}\sigma_{z}  
    +H_{\mbox{\tiny{MS}}}^{\mathrm{ho}}(\omega_{\mathrm{c}}, 0)
    +\mathrm{g}\sqrt{2\hbar\omega_{\mathrm{c}}}\sigma_{x}\hat{x},
    \label{eq:qr-hamiltonian'} 
\end{equation}
where
$H_{\mbox{\tiny{MS}}}^{\mathrm{ho}}(\omega_{\mathrm{c}}, 0)=\frac{1}{2}\hat{p}^{2}+\frac{\omega_{\mathrm{c}}^{2}}{2}\hat{x}^{2}$.

We consider a unitary transformation of the quantum Rabi
Hamiltonian $H_{\mbox{\tiny{QR}}}$.  
For the unitary operator
$U_{0}=\frac{1}{\sqrt{2}}
\begin{pmatrix}
  1 & -1 \\
  1 & 1
\end{pmatrix}
$,
the transformed Hamiltonian is expressed as
\begin{equation}
U_{0}^{*}H_{\mbox{\tiny{QR}}}U_{0}
=
-\, \frac{\hbar\omega_{\mathrm{a}}}{2}\sigma_{x}
+H_{\mbox{\tiny{MS}}}^{\mathrm{ho}}(\omega_{\mathrm{c}}, 0)
+\mathrm{g}\sqrt{2\hbar\omega_{\mathrm{c}}}\sigma_{z}\hat{x}.
    \label{eq:tr_qr-hamiltonian'}
\end{equation}
Consequently, the transformed Hamiltonian $U_{0}^{*}H_{\mbox{\tiny{QR}}}U_{0}$
is the single-mode version of the spin-boson Hamiltonian presented in Eq.(1.4) of Ref.\cite{leg87}.
Following the physical interpretation in Ref.\cite{leg87},
in the $\sigma_{z}$-representation,
the $X$-gate $\sigma_{x}$ acts as the tunneling matrix
exchanging the two states.
Furthermore, the coupling strength $\mathrm{g}$ serves as a parameter
representing the distance between the two minima
of the symmetric double-well potential.
Although the harmonic oscillators of the spin-boson model in Ref.\cite{leg87}
consist of infinitely many modes (representing a heat bath),
our transformed Hamiltonian $U_{0}^{*}H_{\mbox{\tiny{QR}}}U_{0}$
interacts with only a single-mode harmonic oscillator.
Therefore, we assign it a distinct physical interpretation.
We consider the separation of the symmetric wells in a double-well potential  \cite{razavy03}.
As noted in Refs.\cite{leg87,nt99}, a symmetric double-well potential can approximate a qubit
(i.e., a two-level system), a concept essential for implementing qubits in superconducting circuits.
Here, we introduce an approximation in which these roles are reversed.
Although the two-level system is conventionally assumed to be coupled
to a heat bath described by a Bose field in the Caldeira-Leggett model  \cite{cl81, cl83},
we reexamine the Hamiltonian given by Eq.(\ref{eq:tr_qr-hamiltonian'})
in light of Coleman's idea \cite{col77,CC77,col85}.

As reported in Refs.\cite{hs10,hs11,hk13,hk14},
the wavefunction of a Schr\"{o}dinger or Dirac particle may acquire
a phase factor at the barrier boundary during tunneling.
We show below that a similar phenomenon occurs in the quantum Rabi model.
To do this, we introduce two unitary operators. 
We define the unitary operator $U_{1}$ by
$U_{1}=
\begin{pmatrix}
  \exp\left[ -i\frac{\pi}{2}a^{\dagger}a\right] & 0 \\
    0 & \exp\left[ -i\frac{\pi}{2}a^{\dagger}a\right] 
\end{pmatrix}$.
It is noteworthy that the transformation by the diagonal component operator performs a type of Fourier transform:
\begin{align}
\begin{cases}
    e^{i\frac{\pi}{2}a^{\dagger}a}\hat{x}e^{-i\frac{\pi}{2}a^{\dagger}a}
    =\frac{1}{\omega_{\mathrm{c}}}\hat{p}, \\
    e^{i\frac{\pi}{2}a^{\dagger}a}\hat{p}e^{-i\frac{\pi}{2}a^{\dagger}a}
    =-\omega_{\mathrm{c}}\hat{x}.
\end{cases}
\label{eq:F-transformation_1'}
\end{align}
We define the unitary operator $U_{\phi}$ by  
$U_{\phi}=
\begin{pmatrix}
  e^{-i\frac{\mathrm{g}}{\omega_{\mathrm{c}}}(a^{\dagger}+a)} & 0 \\
  0 & e^{i\frac{\mathrm{g}}{\omega_{\mathrm{c}}}(a^{\dagger}+a)}
\end{pmatrix}$.
Then, our Hamiltonian of the transformed quantum Rabi model is given by
\begin{align*}
\widetilde{H}_{\mbox{\tiny QR}}=(U_{0}U_{1}U_{\phi})^{*}H_{\mbox{\tiny QR}}(U_{0}U_{1}U_{\phi}).
\end{align*}

We first note that $(U_{0}U_{1}U_{\phi})^{*}(\mathbb{C}^{2}\otimes\mathcal{F}_{\mathrm{b}})=\mathbb{C}^{2}\otimes\mathcal{F}_{\mathrm{b}}$,
which is the state space of a qubit and a single-mode Bose quasi-particle. 
As shown in Section 2 of Ref.\cite{hir24}, for the grading operator $N_{\mbox{\tiny{F}}}=-\sigma_{z}$, 
a bosonic state is given by $\mid\downarrow\rangle\!\!\mid\!\!\psi\rangle$
for $\psi\in\mathcal{F}_{\mathrm{b}}$ such that
$N_{\mbox{\tiny{F}}}\mid\downarrow\rangle\!\!\mid\!\!\psi\rangle
=\mid\downarrow\rangle\!\!\mid\!\!\psi\rangle$; thus, the space of bosonic states is
$\mathcal{F}_{\downarrow}\!=\mid\downarrow\rangle\mathcal{F}_{\mathrm{b}}$.
Meanwhile, a fermionic state is given by $\mid\uparrow\rangle\!\!\mid\!\!\psi\rangle$
for $\psi\in\mathcal{F}_{\mathrm{b}}$ such that
$N_{\mbox{\tiny{F}}}\mid\uparrow\rangle\!\!\mid\!\!\psi\rangle
=-\mid\uparrow\rangle\!\!\mid\!\!\psi\rangle$,
and the space of fermionic states is
$\mathcal{F}_{\uparrow}\!=\mid\uparrow\rangle\mathcal{F}_{\mathrm{b}}$.
The total state space is decomposed into the orthogonal sum of the bosonic-state
and fermionic-state spaces; hence, the decomposed state space is unitarily equivalent
to the orthogonal sum of the two single-mode Fock spaces,
\begin{equation}
  \mathbb{C}^{2}\otimes\mathcal{F}_{\mathrm{b}}=
  \mathcal{F}_{\downarrow}\oplus\mathcal{F}_{\uparrow}
  \cong
  \mathcal{F}_{\mathrm{b}}\oplus\mathcal{F}_{\mathrm{b}},  
  \label{eq:orthogonal-decomposition}
\end{equation}
through the correspondences,
$$
\begin{pmatrix}
  c_{1} \\ c_{2}
\end{pmatrix}
\otimes\varphi
=\mid\downarrow\rangle\otimes c_{2}\mid\!\!\varphi\rangle
\oplus\mid\uparrow\rangle\otimes c_{1}\mid\!\!\varphi\rangle
\cong
\begin{pmatrix}
  c_{1}\varphi \\ c_{2}\varphi
\end{pmatrix}
=
\begin{pmatrix}
  \psi_{1} \\ \psi_{2}
\end{pmatrix},
$$
where $c_{1}, c_{2}\in\mathbb{C}$, $\varphi\in\mathcal{F}_{\mathrm{b}}$,
and $\psi_{1}=c_{1}\varphi, \psi_{2}=c_{2}\varphi\in\mathcal{F}_{\mathrm{b}}$.
Then, the spin-annihilation operator $\sigma_{-}=\frac{1}{2}(\sigma_{x}-i\sigma_{y})$
and spin-creation operator $\sigma_{+}=\frac{1}{2}(\sigma_{x}+i\sigma_{y})$
describe particle tunneling between the bosonic-state space
$\mathcal{F}_{\downarrow}$ and the fermionic-state space $\mathcal{F}_{\uparrow}$. 
Furthermore, the two single-mode Fock spaces $\mathcal{F}_{\mathrm{b}}$
on the right-hand side of Eq.(\ref{eq:orthogonal-decomposition}) can be
identified with $L^{2}(-\infty, -\Lambda)$ and $L^{2}(+\Lambda, +\infty)$
for a constant $\Lambda\ge 0$, respectively.
Consequently, we obtain the state space for the wavefunctions,
$L^{2}(-\infty, -\Lambda)\oplus L^{2}(+\Lambda, +\infty)
\cong L^{2}(\mathbb{R}\setminus [-\Lambda, +\Lambda])$,
where the interval $[-\Lambda, +\Lambda]$ effectively acts
as a barrier for the wavefunctions \cite{alb88,hs10,hs11,hk13,hk14}.

Setting the operators $A$ and $B$ as 
\begin{align*}
A&=
\begin{pmatrix}
  H_{\mbox{\tiny{MS}}}^{\mathrm{ho}}(\omega_{\mathrm{c}}, 0) & 0 \\
  0 & H_{\mbox{\tiny{MS}}}^{\mathrm{ho}}(\omega_{\mathrm{c}}, 0)
\end{pmatrix}
-\hbar\frac{\mathrm{g}^{2}}{\omega_{\mathrm{c}}},
\nonumber \\
B&=
\begin{pmatrix}
  0 & e^{i\mathrm{g}\sqrt{\frac{8}{\hbar\omega_{\mathrm{c}}}}\hat{x}} \\
e^{-i\mathrm{g}\sqrt{\frac{8}{\hbar\omega_{\mathrm{c}}}}\hat{x}}  & 0    
\end{pmatrix}
=e^{-i\mathrm{g}\sqrt{\frac{8}{\hbar\omega_{\mathrm{c}}}}\hat{x}}\sigma_{-}+e^{i\mathrm{g}\sqrt{\frac{8}{\hbar\omega_{\mathrm{c}}}}\hat{x}}\sigma_{+},
  \label{eq:expression-B}
\end{align*}
we obtain the expression,
\begin{equation}
  \widetilde{H}_{\mbox{\tiny QR}}=A-\hbar\frac{\omega_{\mathrm{a}}}{2}B.  
\label{eq:trans_qr-Hamiltonian_2}
\end{equation}
The operator $A$ describes the particle's confinement within the potential well of the harmonic oscillator.
The operator $-\hbar\frac{\omega_{\mathrm{a}}}{2}B$ determines the tunneling behavior of the particle through the wells,
thereby governing the tunneling within our two-level-system approximation
by incorporating a weighted phase factor,
\begin{equation}
  \frac{\hbar\omega_{\mathrm{a}}}{2}e^{\pm i\mathrm{g}\sqrt{\frac{8}{\hbar\omega_{\mathrm{c}}}}\hat{x}}. 
  \label{eq:weighted-phase-factor}
\end{equation}
Thus, the particles acquire this phase factor 
as they pass through the barrier, such as a tunnel junction \cite{hs10,hs11,hk13,hk14}. 

In addition to the appearance of the phase factor,
as seen by comparing Eq.(\ref{eq:qr-hamiltonian'}) and Eq.(\ref{eq:trans_qr-Hamiltonian_2}), 
we note that the tunneling does not change the effect of the harmonic-oscillator potential.
The Schr\"{o}dinger operator appearing in the diagonal elements
(i.e., the $(1,1)$- and $(2,2)$-elements) of Eq.(\ref{eq:tr_qr-hamiltonian'})
can be rewritten as
\begin{equation}
  H_{\mbox{\tiny{MS}}}^{\mathrm{ho}}(\omega_{\mathrm{c}}, 0)
  \pm \mathrm{g}\sqrt{2\hbar\omega_{\mathrm{c}}}\hat{x}
  =\frac{1}{2}\hat{p}^{2}
  +\frac{\omega_{\mathrm{c}}^{2}}{2}\left(
  \hat{x}\pm\frac{\mathrm{g}}{\omega_{\mathrm{c}}}\sqrt{\frac{2\hbar}{\omega_{\mathrm{c}}}}
  \right)^{2}
  -\hbar\frac{\mathrm{g}^{2}}{\omega_{\mathrm{c}}}.
  \label{eq:VanHove}
\end{equation}
Therefore, this fact indicates that the tunneling in the two-level-system approximation
using the quantum Rabi model does not change
the shape of the two wells of the harmonic oscillator potential, but
merely shifts the positions of the potential minima
and generates the self-energy $-\hbar\frac{\mathrm{g}^{2}}{\omega_{\mathrm{c}}}$.

In Ref.\cite{hir99}, the author introduced the unitary transformation
analogous to $U_{0}U_{1}U_{\phi}$ solely as a purely mathematical construct;
however, the above description provides a physical interpretation.
This interpretation suggests that the ground-state energy expression
of the spin-boson model \cite{hir99}
reflects the effects of a dilute instanton gas \cite{as10}.

\section{Quantum Rabi Model and Transition to Spontaneous $\mathbb{Z}_{2}$-Symmetry Breaking}
\label{sec:qrm_ssb}

In this section, we introduce two limits that lead the quantum Rabi model
to spontaneous $\mathbb{Z}_{2}$-symmetry breaking.
It is the transformed quantum Rabi model that actually reveals the transition
to this spontaneous $\mathbb{Z}_{2}$-symmetry breaking.

We consider the following two types of limits.
\begin{itemize}
\item[](LMT1) Fixing the frequencies $\omega_{\mathrm{a}}$ and $\omega_{\mathrm{c}}$,
  we treat the coupling strength $\mathrm{g}$ as a parameter
  and take the limit $\mathrm{g}\to\infty$. 
\item[](LMT2) Fixing the frequency $\omega_{\mathrm{c}}$,
  we introduce a variable $r$ with $0\le r\le 1$
  and define the frequency $\omega_{\mathrm{a}}$ and the coupling strength $\mathrm{g}$
  as continuous functions of the variable $r$, i.e., 
  $\omega_{\mathrm{a}}=\omega_{\mathrm{a}}(r)$ and $\mathrm{g}=\mathrm{g}(r)$. 
  These functions satisfy the conditions,
  $\omega_{\mathrm{a}}(r)\ne 0$ for $0\le r<1$, $\omega_{\mathrm{a}}(1)=0$,
  $\mathrm{g}(0)=0$, and $\mathrm{g}(r)\ne 0$ for $0<r\le 1$. 
  We then take the limit $r\to 1$. 
\end{itemize}
We denote by ${\displaystyle \lim_{\mbox{\tiny{SB}}}}$ 
the limit given by (LMT1) or (LMT2).
In the limit (LMT1), we recall that the parameter $\mathrm{g}$ represents the distance
between the two minima of the symmetrical wells \cite{leg87} (also see Eq.(\ref{eq:VanHove})). 
Therefore, the adiabatic approximation becomes mathematically valid in the strong-coupling limit
$\mathrm{g}\to\infty$ \cite{hir20},
as the tunneling from one minimum to another is virtually impossible
for sufficiently large $\mathrm{g}$.
Exploiting this, we can allow the frequency $\omega_{\mathrm{a}}$ to assume the role of the coupling strength $\mathrm{g}$,
leading to the limit (LMT2).
This approach was first employed experimentally by Cai \textit{et al}. to realize the transition
using an ion-trap simulator \cite{cai22}.
While the transition to spontaneous $\mathbb{Z}_{2}$-symmetry breaking is obtainable
in both limits for the quantum Rabi model,
it does not occur in (LMT1) when the so-called $A^{2}$--term (i.e., mass term)
$\hbar C\mathrm{g}^{2}(a+a^{\dagger})^{2}$ is added to the Hamiltonian,
where $C$ is an arbitrary non-zero constant.
This is due to the divergence of the free boson Hamiltonian after renormalization
via the Hopfield-Bogoliubov transformation \cite{hir20}.
In contrast, the transition remains achievable in (LMT2) \cite{hir24}.

We define the Hamiltonian of the renormalized quantum Rabi model as:
\begin{align*}
H_{\mbox{\tiny{QR}}}^{\mathrm{ren}}=
H_{\mbox{\tiny{QR}}}+\hbar\frac{\mathrm{g}^{2}}{\omega_{\mathrm{c}}}.
\end{align*}
Correspondingly, its unitary transformation is given by
\begin{align*}
\widetilde{H}_{\mbox{\tiny{QR}}}^{\mathrm{ren}}=
(U_{0}U_{1}U_{\phi})^{*}H_{\mbox{\tiny{QR}}}^{\mathrm{ren}}(U_{0}U_{1}U_{\phi})
=A+\hbar\frac{\mathrm{g}^{2}}{\omega_{\mathrm{c}}}-\hbar\frac{\omega_{\mathrm{a}}}{2}B.
\end{align*}
It is obvious that the operator $-\frac{\hbar\omega_{\mathrm{a}}}{2}B$ converges to $0$
in (LMT2).
Meanwhile, regarding the expectation value of this operator for any state
$\mid\!\!\sharp\rangle\!\!\mid\!\!\psi\rangle$,
the weighted phase factors in Eq.(\ref{eq:weighted-phase-factor})
--- appearing in the off-diagonal elements ---
yield individual vibration integrals of the form
$\int_{-\infty}^{+\infty}e^{\pm i\mathrm{g}\sqrt{\frac{8}{\hbar\omega_{\mathrm{c}}}}x}\mid\!\!\psi(x)\!\!\mid^{2}dx$.
These vibration integrals converge to $0$ in (LMT1) \cite{hir15}
since the vibrations become increasingly rapid as $\mathrm{g}\to\infty$.
More generally, for arbitrary wavefunctions $\varphi(x)$ and $\psi(x)$,
the integrals $\int_{-\infty}^{+\infty}\varphi^{*}(x)e^{\pm i\mathrm{g}\sqrt{\frac{8}{\hbar\omega_{\mathrm{c}}}}x}\psi(x)dx$
vanish in (LMT1) by the Riemann-Lebesgue lemma. 
We mathematically justify this heuristic argument below. 
By combining these convergences with Eq.(\ref{eq:trans_qr-Hamiltonian_2}),
and following the methods in Refs.\cite{hir15, hir17, hir24},
we can show that the difference between $H_{\mbox{\tiny{QM}}}^{\mathrm{ren}}$ and
$(U_{0}U_{1}U_{\phi})H_{\mathrm{b}}(U_{0}U_{1}U_{\phi})^{*}$
converges to $0$ in the norm resolvent sense (see the definition in Ref.\cite{rs1}, p.284):
\begin{equation}
\lim_{\mbox{\tiny{SB}}}\Bigl\|
\left(\widetilde{H}_{\mbox{\tiny{QM}}}^{\mathrm{ren}}-z\right)^{-1}
-\left(H_{\mathrm{b}}-z\right)^{-1}
\Bigr\|_{\mathrm{op}}=0
\label{eq:norm-resolvent-convergence}
\end{equation}
for all $z\in \mathbb{C}$ with $\Im z\ne 0$,  
where $\|\quad\|_{\mathrm{op}}$ denotes the operator norm,
and $H_{\mathrm{b}}$ stands for the free-boson Hamiltonian$1\otimes\hbar\omega_{\mathrm{c}}\left(a^{\dagger}a+\frac{1}{2}\right)$. 
The norm resolvent convergence implies the convergence of the energy levels
(see Theorems VIII 23 and VIII 24 of Ref.\cite{rs1}).  
We denote the limit in Eq.(\ref{eq:norm-resolvent-convergence}) by
\begin{equation}
\lim_{\mbox{\tiny{SB}}}\widetilde{H}_{\mbox{\tiny{QR}}}^{\mathrm{ren}}=
H_{\mathrm{b}} 
\,\,\, \mbox{or}\,\,\,   
  \lim_{\mbox{\tiny{SB}}}H_{\mbox{\tiny{QM}}}^{\mathrm{ren}}
  \cong H_{\mathrm{b}}.
  \label{eq:norm-resolvent-convergence'}
\end{equation}

\begin{figure}[h]
\centering
\includegraphics[width=0.45\textwidth]{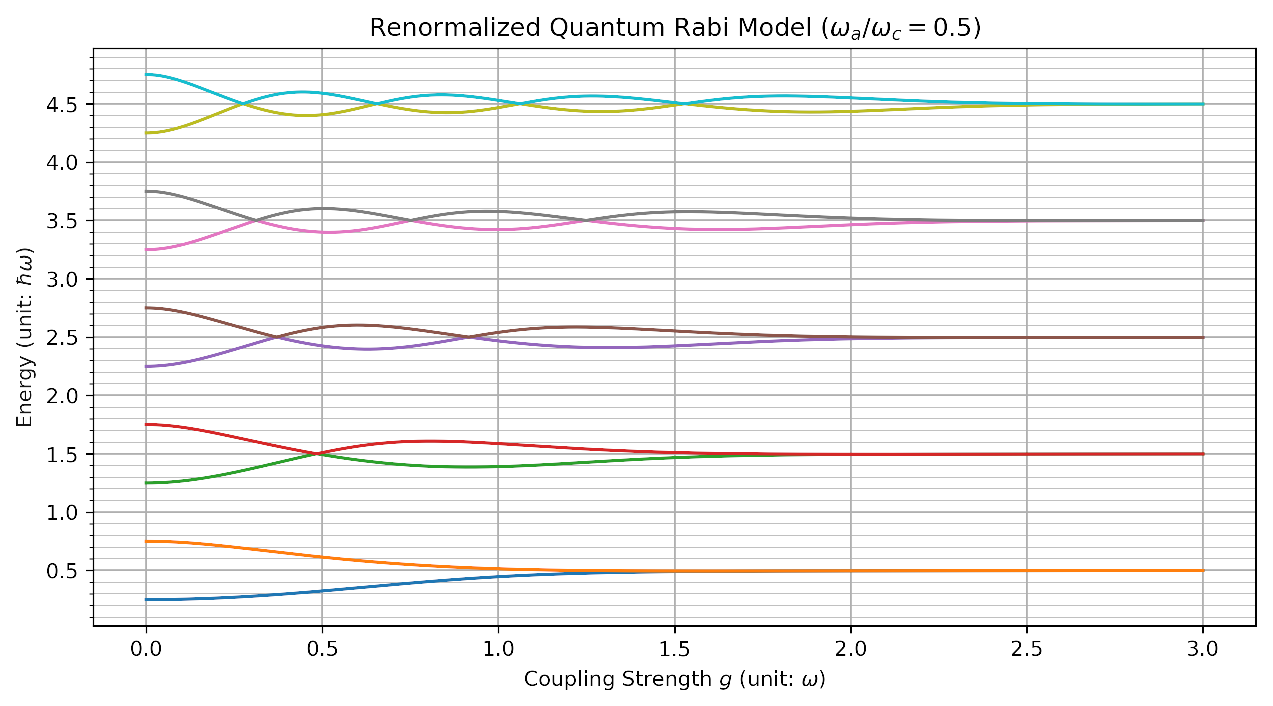}\qquad
\includegraphics[width=0.45\textwidth]{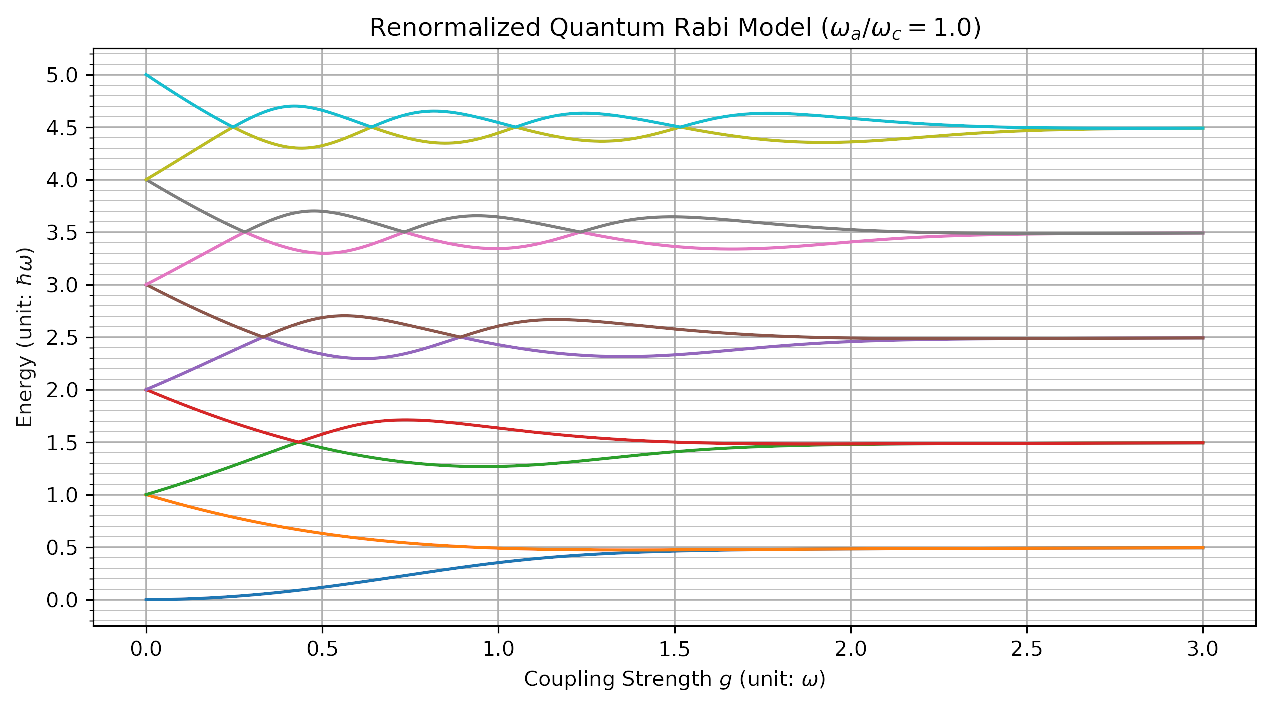}
\caption{The first ten energy levels of the renormalized quantum Rabi model in the limit (LMT1).
  The qubit transition frequency, the frequency of the single-mode boson,
  and the coupling strength between them are denoted by $\omega_{\mathrm{a}}$, $\omega_{\mathrm{c}}$,
  and $\mathrm{g}$, respectively.
  For a constant frequency $\omega$,
  the left panel shows the energy levels for $\omega_{\mathrm{a}}=0.5\omega$ and $\omega_{\mathrm{c}}=\omega$,
  while the right panel displays the case where $\omega_{\mathrm{a}}=\omega_{\mathrm{c}}=\omega$.
  At $\mathrm{g}=0$, the $\mathcal{N}=2$ SUSY is observed in the right graph,
  whereas it is absent in the left panel.} 
\label{fig:renqr_energy_1}
\end{figure}

\begin{figure}[h]
\centering
\includegraphics[width=0.45\textwidth]{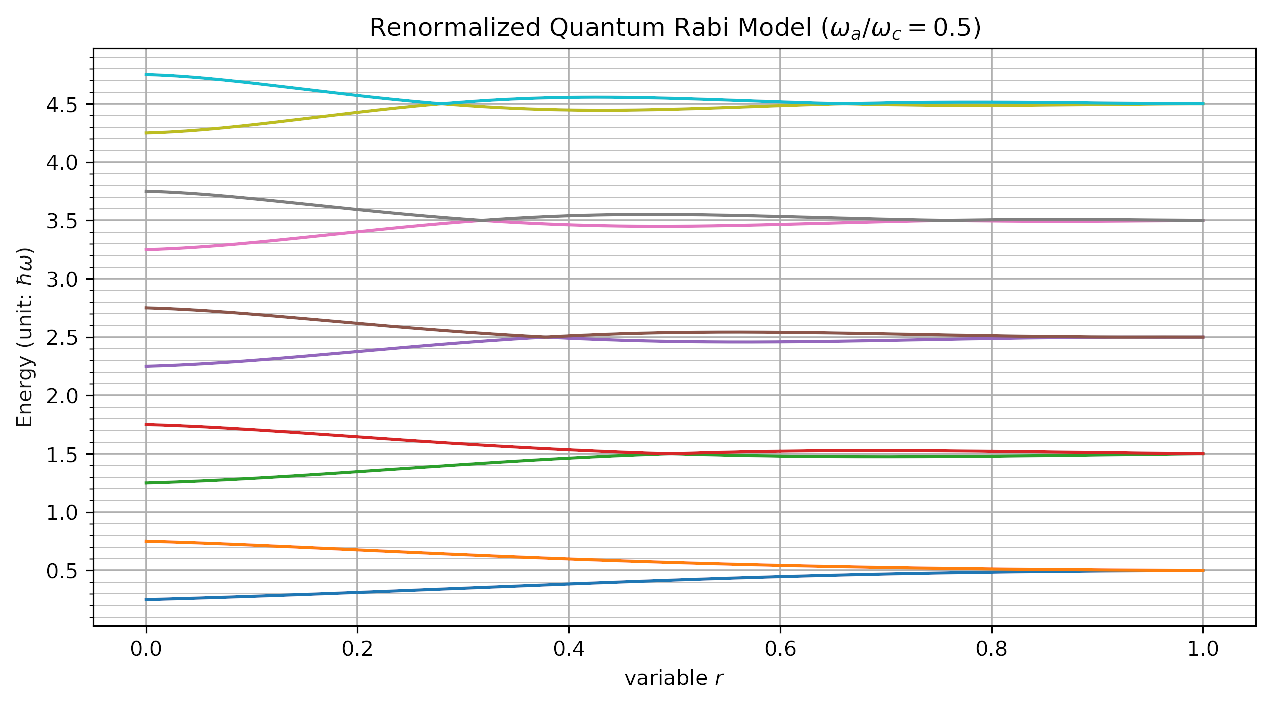}\qquad
\includegraphics[width=0.45\textwidth]{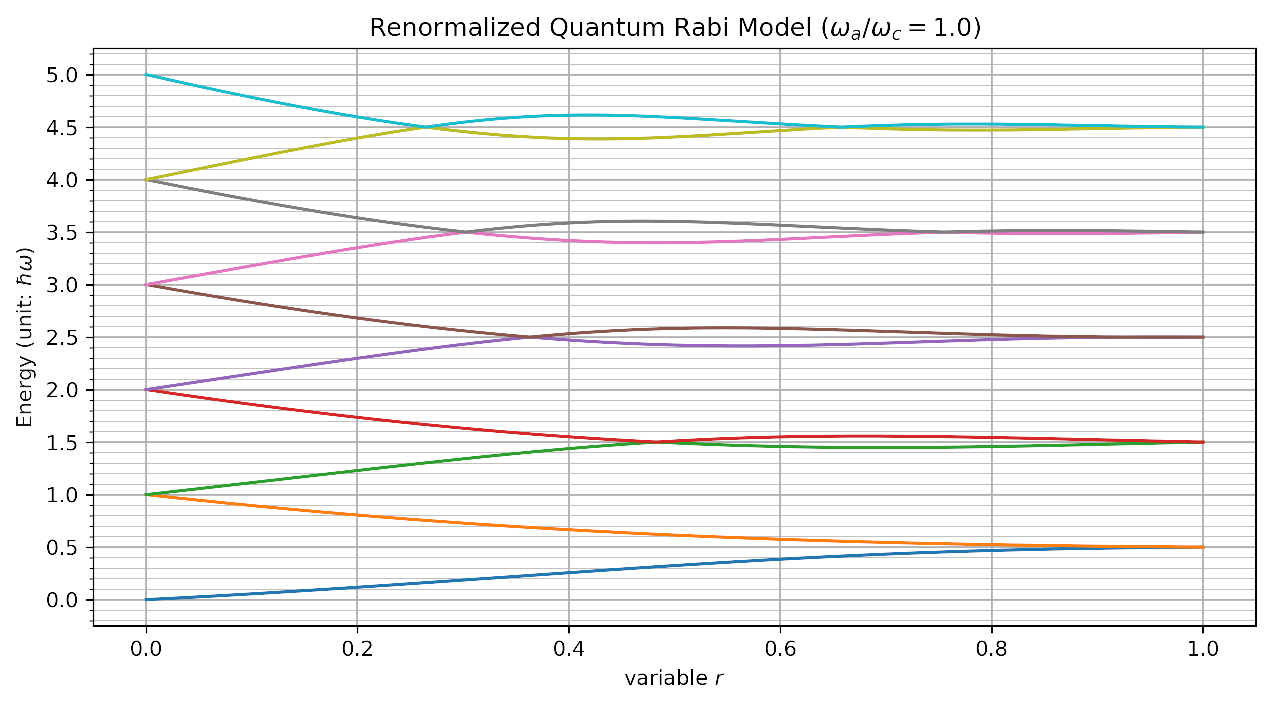}
\caption{The first ten energy levels of the renormalized quantum Rabi model in the limit (LMT2). 
  Here, the qubit transition frequency, the frequency of the single-mode boson,
  and the coupling strength between the qubit and the boson are denoted by
  $\omega_{\mathrm{a}}(r)$, $\omega_{\mathrm{c}}$, and $\mathrm{g}(r)$ 
  ($0\le r\le 1$), respectively.
  For a constant frequency $\omega$,
  the left panel shows the energy levels for $\omega_{\mathrm{a}}(0)=0.5\omega$ and $\omega_{\mathrm{c}}=\omega$,
  while the right panel shows them for $\omega_{\mathrm{a}}(0)=\omega_{\mathrm{c}}=\omega$.
  At $r=0$ (i.e., $\mathrm{g}(0)=0$), the $\mathcal{N}=2$ SUSY can be found in the right panel,
  but not in the left panel.} 
\label{fig:renqr_energy_1_r}
\end{figure}

For the parity operator $P$, we define the transformed parity operator $\widetilde{P}$
by $\widetilde{P}=(U_{0}U_{1}U_{\phi})^{*}P(U_{0}U_{1}U_{\phi})$,
and we have 
$\widetilde{P}=-\sigma_{x}(-1)^{a^{\dagger}a}$.
This is the parity operator for the tunneling and the boson number.
Eq.(\ref{eq:parity-symmetry}) ensures the parity symmetry of the transformed quantum Rabi Hamiltonian:
\begin{equation}
  \bigl[\widetilde{H}_{\mbox{\tiny{QR}}},\widetilde{P}\bigr]=\bigl[\widetilde{H}^{\mathrm{ren}}_{\mbox{\tiny{QR}}},\widetilde{P}\bigr]=0.
\label{eq:parity-symmetry'}
\end{equation}
Meanwhile, it is obvious that the $\mathbb{Z}_{2}$-symmetry,
\begin{equation}
  [-\sigma_{x}, H_{\mathrm{b}}]=0,
  \label{eq:Z_2-symmetry}
\end{equation}
holds.
This implies the following $\mathbb{Z}_{2}$-symmetry,
\begin{equation}
[\widetilde{P} , H_{\mathrm{b}}]=0
  \label{eq:another_Z_2-symmetry}
\end{equation}
since $[a^{\dagger}a, H_{\mathrm{b}}]=0$.
Moreover, the Hamiltonian $H_{\mathrm{b}}$ possesses degenerate ground states,
$\mid\downarrow\rangle\!\!\mid\!\!\!0\rangle$
and
$\mid\uparrow\rangle\!\!\mid\!\!\!0\rangle$. 
This surely implies the spontaneous $\mathbb{Z}_{2}$-symmetry breaking.
Thus, the limit in the sense of (LMT1) or (LMT2)
drives the renormalized Hamiltonian $H_{\mbox{\tiny{QR}}}^{\mathrm{ren}}$ toward
spontaneous $\mathbb{Z}_{2}$-symmetry breaking
(see Figs.\ref{fig:renqr_energy_1} and \ref{fig:renqr_energy_1_r}).
Eqs.(\ref{eq:Z_2-symmetry}) and (\ref{eq:another_Z_2-symmetry})
then characterize the difference in $\mathbb{Z}_{2}$-symmetry,
reflecting the emergence of the single-mode boson (i.e., Bose quasi-particle) penetration:
\begin{equation}
  \begin{cases}
[\widetilde{P} , \widetilde{H}_{\mbox{\tiny{QR}}}^{\mathrm{ren}}]=0=[-\sigma_{x} , \widetilde{H}_{\mbox{\tiny{QR}}}^{\mathrm{ren}}]
& \mbox{in LMT1 or LMT2}, \\
[\widetilde{P} , \widetilde{H}_{\mbox{\tiny{QR}}}^{\mathrm{ren}}]=0\ne [-\sigma_{x} , \widetilde{H}_{\mbox{\tiny{QR}}}^{\mathrm{ren}}]
& \mbox{for $\frac{\mathrm{g}}{\omega_{\mathrm{a}}}<\infty$}. \\
  \end{cases}
  \label{eq:change_Z_2-symmetry}
\end{equation}
The first commutation relation in Eq.(\ref{eq:change_Z_2-symmetry})
indicates that the tunneling between the bosonic states $\mathcal{F}_{\downarrow}$
and the fermionic states $\mathcal{F}_{\uparrow}$ is independent of
the penetration of the single-mode boson (i.e., Bose quasi-particle).
Once the tunneling is activated, it breaks one of the two $\mathbb{Z}_{2}$-symmetries,
as evidenced by $[-\sigma_{x} , \widetilde{H}_{\mbox{\tiny{QR}}}^{\mathrm{ren}}]\ne 0$ in Eq.(\ref{eq:change_Z_2-symmetry}).
In contrast, the other $\mathbb{Z}_{2}$-symmetry remains intact for all parameters, $\omega_{\mathrm{a}}, \omega_{\mathrm{c}}$, and $\mathrm{g}$,  
as $[\widetilde{P} , \widetilde{H}_{\mbox{\tiny{QR}}}^{\mathrm{ren}}]=0$ in Eq.(\ref{eq:change_Z_2-symmetry}).
This implies that the tunneling is assisted by the penetration of the single-mode boson.

We now tune the frequencies so that the strict condition, $\omega_{\mathrm{a}}=\omega_{\mathrm{c}}$,
is satisfied.
Since the quantum Rabi Hamiltonian $H_{\mbox{\tiny{QR}}}$ with $\mathrm{g}=0$
corresponds to the so-called Witten Hamiltonian \cite{wit81,wit82},
SUSY is obtained under this resonance condition, $\omega_{\mathrm{a}}=\omega_{\mathrm{c}}$
(see the right panels of Figs.\ref{fig:renqr_energy_1} and \ref{fig:renqr_energy_1_r}).
It is worthy to note that the tuning of $\omega_{\mathrm{a}}=\omega_{\mathrm{c}}$ is straightforward in certain quantum simulators.

The anisotropic quantum Rabi model \cite{tomka14,shen14,xie14,zhang15,wang19} describes
the interaction between a qubit and a single-mode boson in a cavity.
The interaction is defined by interpolating between the quantum Rabi model and the Jaynes-Cummings model,
which breaks the balance between the rotating terms and the counter-rotating terms of
the linear interaction.
In the strong-coupling limit ($\mathrm{g}\to\infty$),
the anisotropic quantum Rabi model undergoes a transition to
spontaneous $\mathbb{Z}_{2}$-symmetry breaking that is stable
against variations in the frequencies $\omega_{\mathrm{a}}$ and $\omega_{\mathrm{c}}$,
as well as the imbalance between rotating and counter-rotating terms \cite{hhl25}.
Therefore, under the condition $\omega_{\mathrm{a}}=\omega_{\mathrm{c}}$,
the anisotropic quantum Rabi model exhibits a transition from the $\mathcal{N}=2$ SUSY
to its spontaneous breaking.
However, this imbalance causes a mass reduction \cite{hhl25},
as the tunneling effect induces momentum penetration (or, alternatively,
position penetration according to Eq.(\ref{eq:F-transformation_1'})).
Thus, it is worth studying another SUSY and its spontaneous breaking
within the context of the anisotropic quantum Rabi model \cite{kafuri24}.

\section{Occurrence of Polyakov's Energy Bifurcation}
\label{sec:polyakov}

In this section, we demonstrate the restoration of the spontaneously broken $\mathbb{Z}_{2}$-symmetry
in the (renormalized) quantum Rabi model, following Polyakov's assertion \cite{pol77}.

\begin{figure}[h]
\centering
\includegraphics[width=0.45\textwidth]{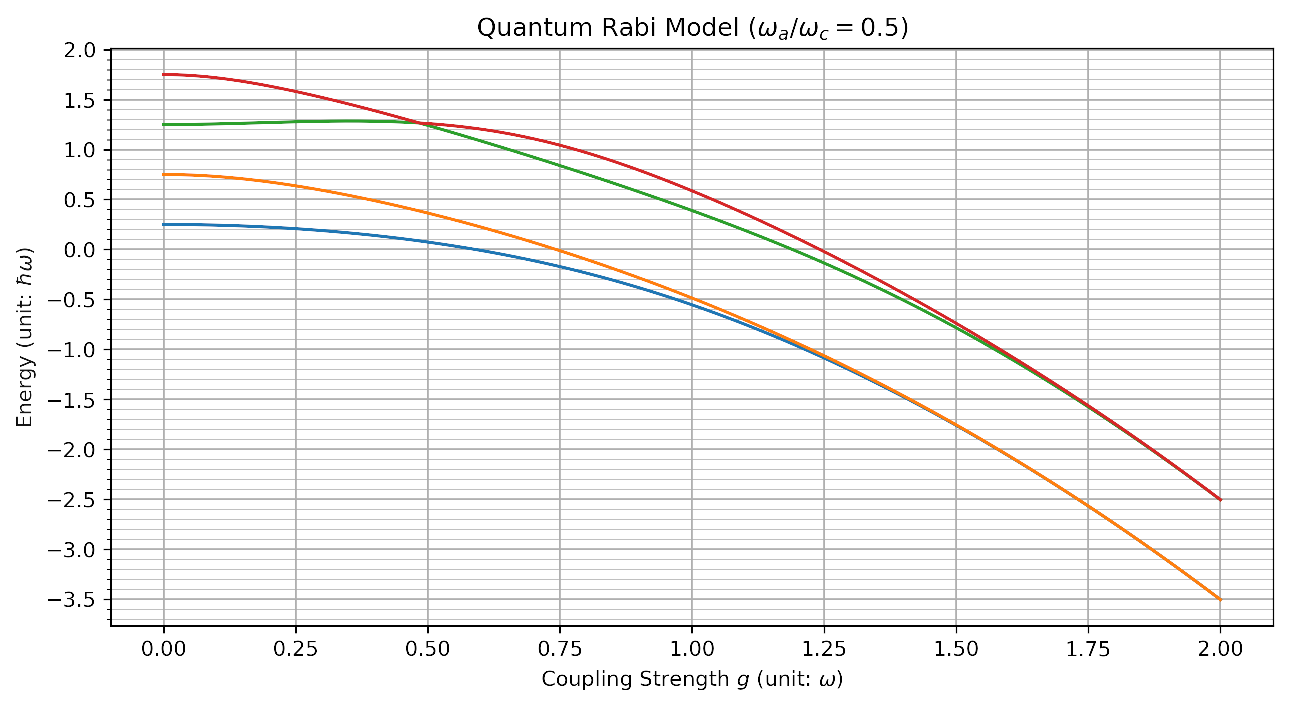}\qquad 
\includegraphics[width=0.45\textwidth]{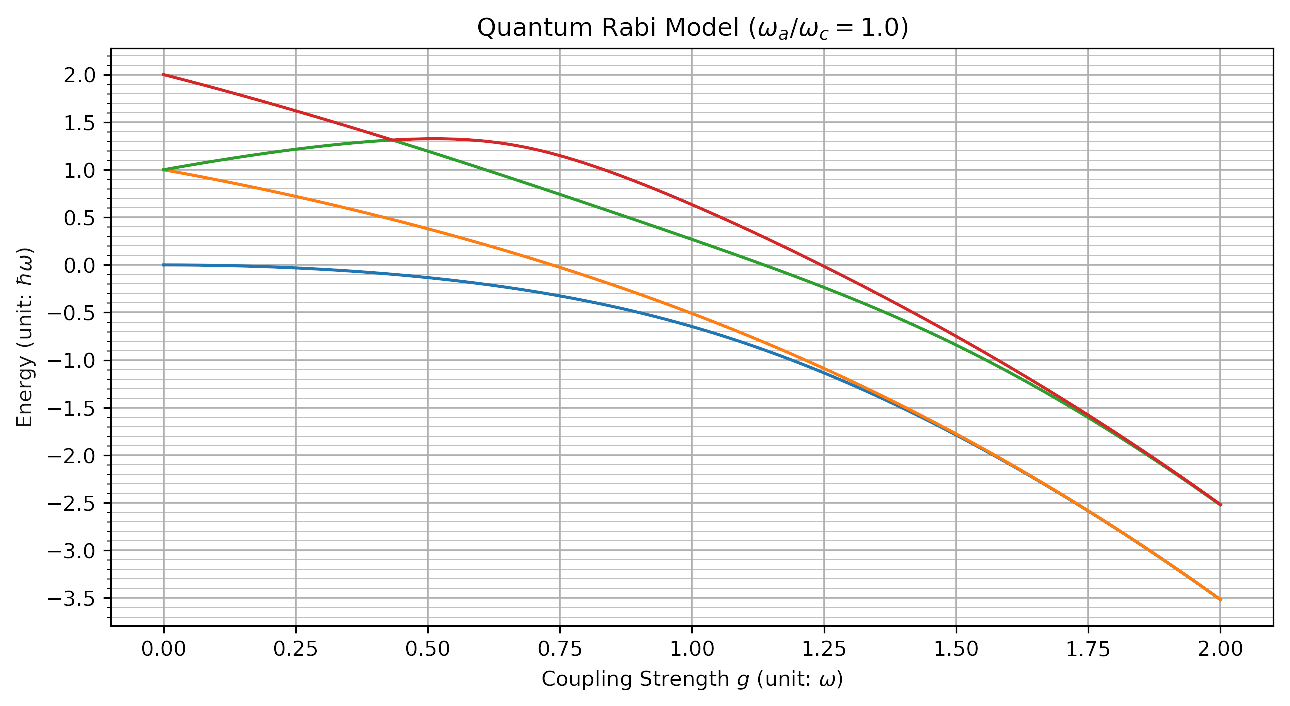}
\caption{The four lowest energy levels of the quantum Rabi model from (LMT1).
  Here, the qubit-transition frequency, the single-mode boson frequency,
  and the coupling strength between the qubit and the boson are denoted
  by $\omega_{\mathrm{a}}$, $\omega_{\mathrm{c}}$,   and $\mathrm{g}$, respectively.
  For a constant frequency $\omega$,
  the left panel shows the energy levels in the cases
  $\omega_{\mathrm{a}}=0.5\omega$ and $\omega_{\mathrm{c}}=\omega$,
  while the right panel shows the case $\omega_{\mathrm{a}}=\omega_{\mathrm{c}}=\omega$.
  At $\mathrm{g}=0$, we find the $\mathcal{N}=2$ SUSY in the right panel,
  but not in the left one.} 
\label{fig:qr_energy_2}
\end{figure}

\begin{figure}[h]
\centering
\includegraphics[width=0.45\textwidth]{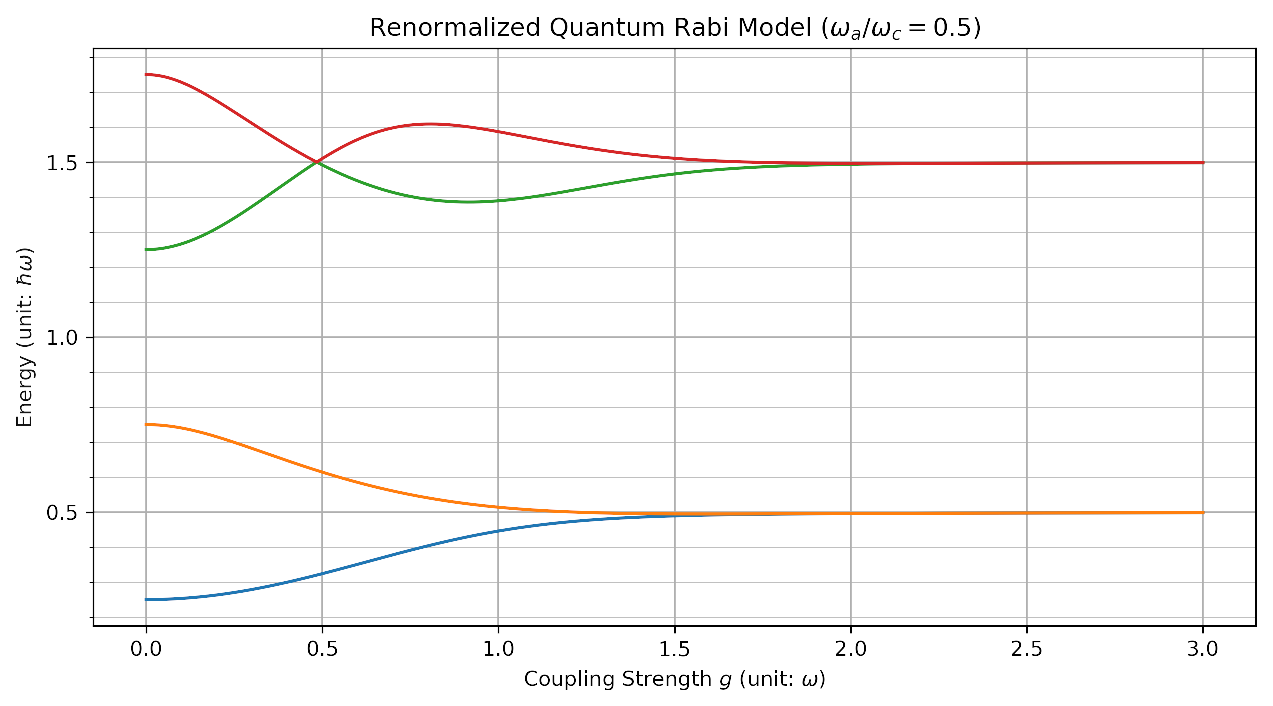}\qquad
\includegraphics[width=0.45\textwidth]{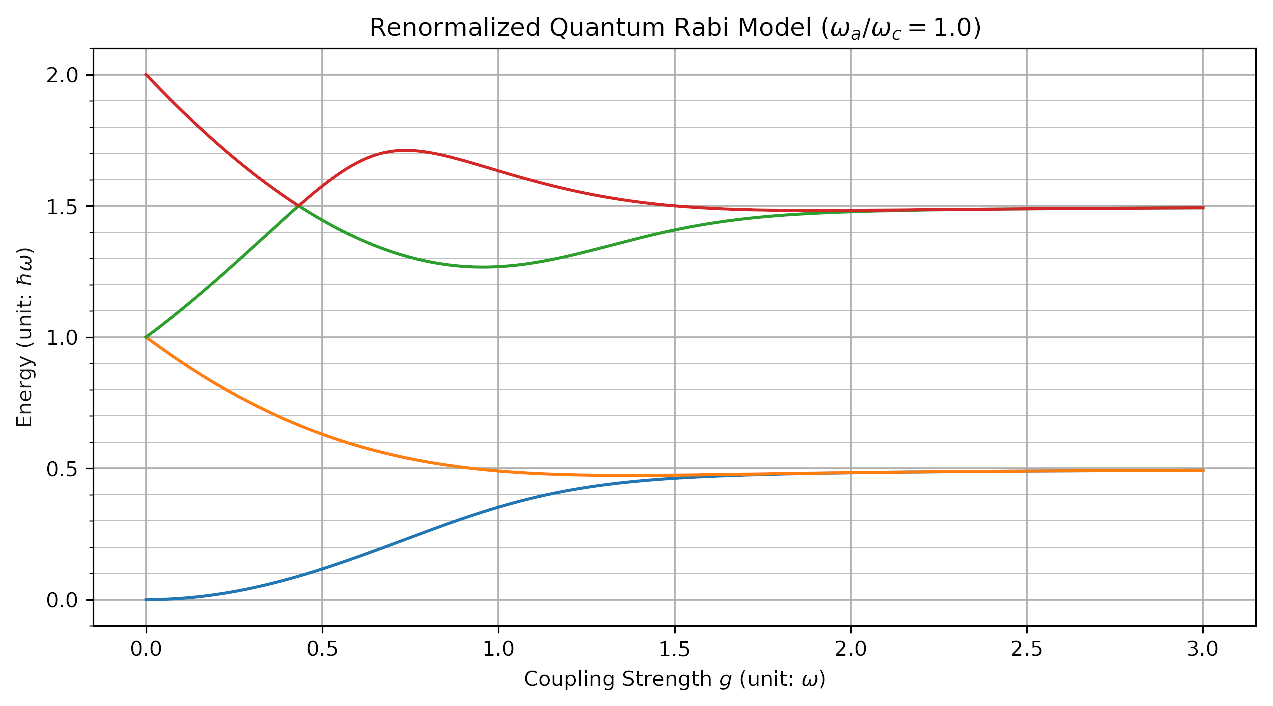}
\caption{The first four energy levels of the renormalized quantum Rabi model in (LMT1).
  Here, $\omega_{\mathrm{a}}$, $\omega_{\mathrm{c}}$, and $\mathrm{g}$ denote
  the qubit-transition frequency, the frequency of the single-mode boson,
  and the coupling strength between them, respectively.
  For a constant frequency $\omega$,
  the left panel shows the energy levels for $\omega_{\mathrm{a}}=0.5\omega$
  and $\omega_{\mathrm{c}}=\omega$,
  while the right panel shows the case where $\omega_{\mathrm{a}}=\omega_{\mathrm{c}}=\omega$.
  At $\mathrm{g}=0$, the $\mathcal{N}=2$ SUSY is observed in the right panel, but not in the left.} 
\label{fig:renqr_energy}
\end{figure}

\begin{figure}[h]
\centering
\includegraphics[width=0.45\textwidth]{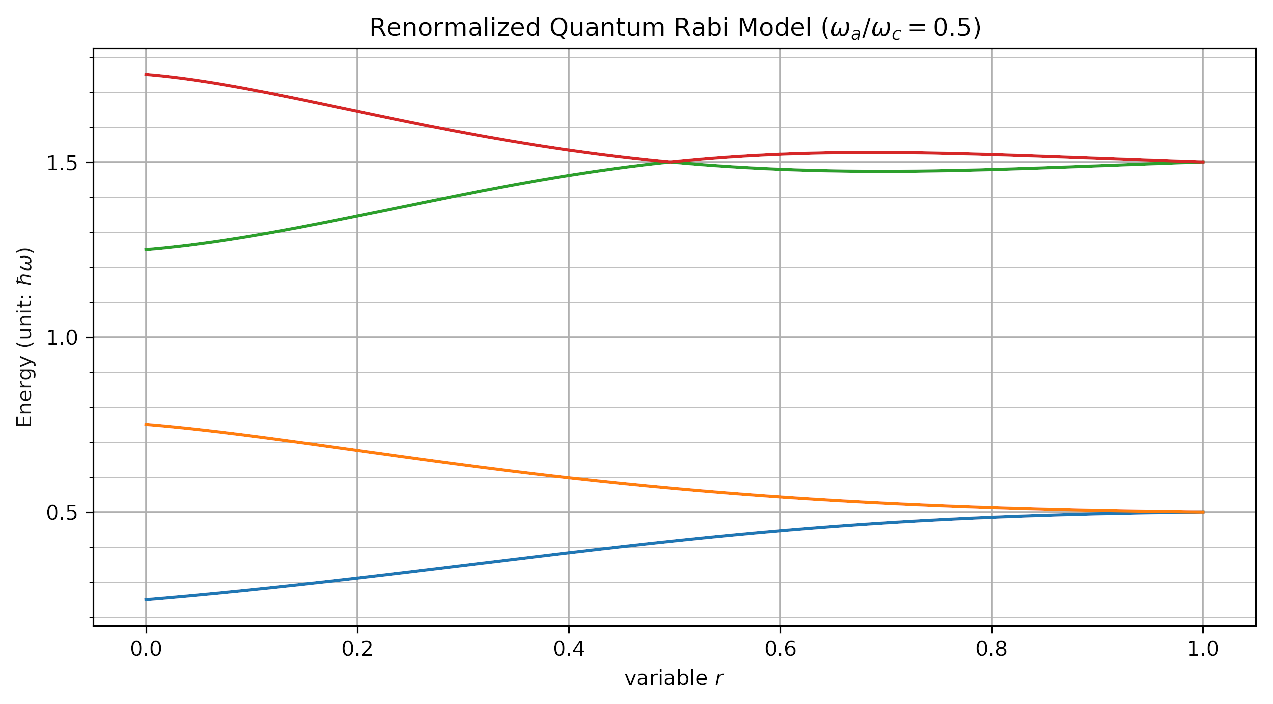}\qquad
\includegraphics[width=0.45\textwidth]{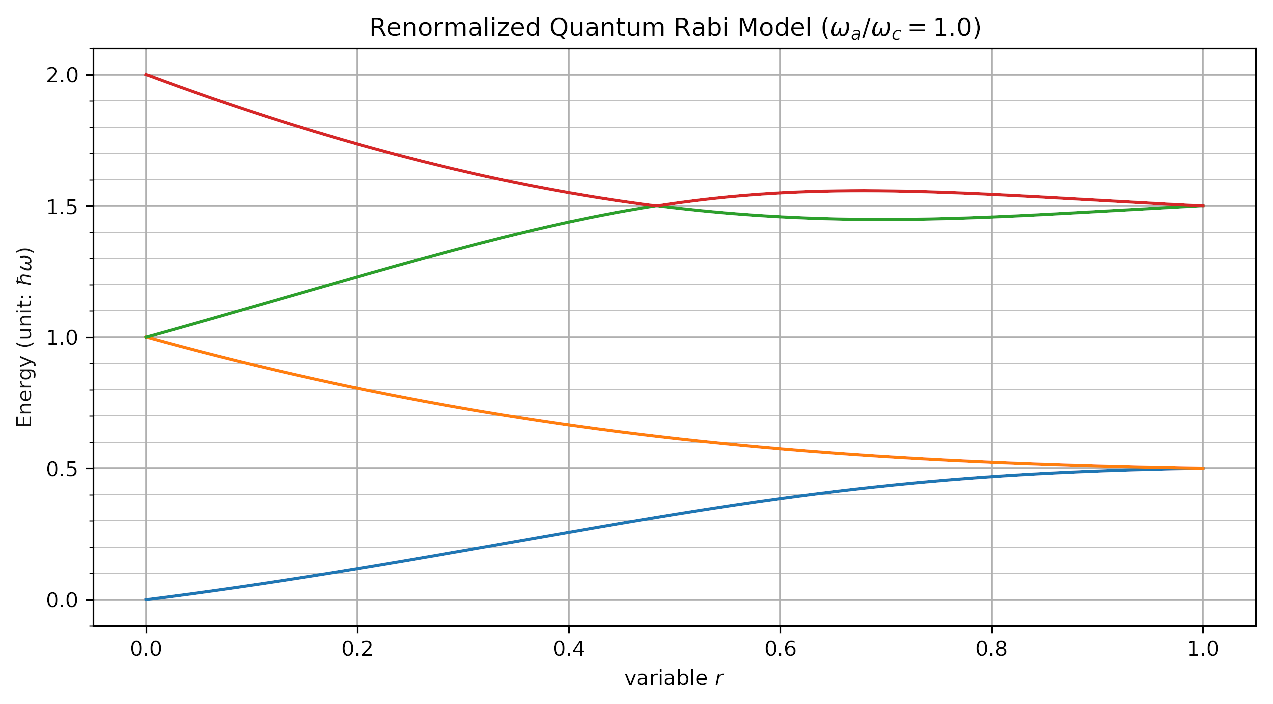}
\caption{Energy levels of the renormalized quantum Rabi model in (LMT2).
  The first four energy levels are shown.
  Here, $\omega_{\mathrm{a}}(r)$, $\omega_{\mathrm{c}}$, and $\mathrm{g}(r)$
  denote the qubit-transition frequency, the frequency of the single-mode boson, and the coupling strength, respectively.
  For a constant frequency $\omega$,
  the left panel shows the energy levels for $\omega_{\mathrm{a}}(0)=0.5\omega$ and $\omega_{\mathrm{c}}=\omega$,
  while the right graph shows the case where $\omega_{\mathrm{a}}(0)=\omega_{\mathrm{c}}=\omega$.
  At $r=0$ (i.e., $\mathrm{g}(0)=0$), we can find $\mathcal{N}=2$ SUSY in the right panel,
  but not in the left one.} 
\label{fig:renqr_energy_r}
\end{figure}

Let us view this recovery as being in the opposite direction to the transition
to spontaneous $\mathbb{Z}_{2}$-symmetry breaking.
When this spontaneous breaking occurs, the renormalized, transformed quantum Rabi Hamiltonian
$\widetilde{H}_{\mbox{\tiny{QR}}}^{\mathrm{ren}}$
no longer possesses a tunneling interaction between the subspaces of bosonic and fermionic states,
as shown in Eq.(\ref{eq:norm-resolvent-convergence'});
consequently, it features degenerate ground states at $\frac{\mathrm{g}}{\omega_{\mathrm{a}}}=\infty$. 
Once $\frac{\mathrm{g}}{\omega_{\mathrm{a}}}$ becomes finite (i.e., $\frac{\mathrm{g}}{\omega_{\mathrm{a}}}<\infty$),
the tunneling interaction emerges, and the Hamiltonian becomes to the full transformed quantum Rabi Hamiltonian.
The (transformed) quantum Rabi Hamiltonian has a unique ground state \cite{hh14},
where the $\mathbb{Z}_{2}$-symmetry is restored from spontaneous breaking.
Thus, the degeneracy is lifted, and the degenerate ground state splits
into the ground state and the first excited state (see, for instance,
Figs.\ref{fig:qr_energy_2}, \ref{fig:renqr_energy}, and \ref{fig:renqr_energy_r}). 
We show below that this phenomenon originates from Polyakov's statement. 

Thanks to the parity symmetry given by Eq.(\ref{eq:parity-symmetry'}),
the total state space $\mathbb{C}^{2}\otimes\mathcal{F}_{\mathrm{b}}$ can be decomposed into the direct sum $\mathcal{F}_{+}\oplus\mathcal{F}_{-}$.
Here, $\mathcal{F}_{\pm}$ are the mutually orthogonal eigenspaces
of the transformed parity operator $\widetilde{P}$ satisfying 
$\widetilde{P}\mathcal{F}_{\pm}=\mp\mathcal{F}_{\pm}$.
In other words, the eigenvalue equation $\widetilde{P}\hspace*{-1.2mm}\mid\hspace*{-1.2mm}\psi_{\pm}\rangle=\mp\hspace*{-1.2mm}\mid\hspace*{-1.2mm}\psi_{\pm}\rangle$
holds for every $\mid\hspace*{-1.2mm}\psi_{\pm}\rangle\in\mathcal{F}_{\pm}$.
Correspondingly, the symmetry in Eq.(\ref{eq:parity-symmetry'}) induces the Hamiltonian decomposition,
$\widetilde{H}_{\mbox{\tiny{QR}}}=\widetilde{H}_{+}+\widetilde{H}_{-}$,
where each $\widetilde{H}_{\pm}$ acts on the domain 
$\mathrm{Dom}(\widetilde{H}_{\pm})\subset\mathcal{F}_{\pm}$
and
leaves the subspace $\mathcal{F}_{\pm}$ invariant, i.e.,
$\widetilde{H}_{\pm}\mathrm{Dom}(\widetilde{H}_{\pm})\subset\mathcal{F}_{\pm}$.
We denote this decomposition by
\begin{equation}
  \widetilde{H}_{\mbox{\tiny{QR}}}=\widetilde{H}_{+}\oplus\widetilde{H}_{-}.
  \label{eq:decomposition_0}
\end{equation}
Similarly to Eq.(2.18) of Ref.\cite{hir99}
or Eqs.(3.98) and (3.99) of Ref.\cite{razavy03}, 
we define the states $\widetilde{\Omega}_{\pm}
=(\mid\uparrow\rangle\!\!\mid\!\!0\rangle\pm\!\mid\downarrow\rangle\!\!\mid\!\!0\rangle)/\sqrt{2}$
and $\Omega_{\pm}=(U_{0}U_{1}U_{\phi})\widetilde{\Omega}_{\pm}$.
It follows that $\widetilde{\Omega}_{\pm}\in\mathcal{F}_{\pm}$, i.e.,
$\widetilde{P}\widetilde{\Omega}_{\pm}=\mp\widetilde{\Omega}_{\pm}$.
Based on the decomposition given by Eq.(\ref{eq:decomposition_0}), we have
\begin{align*}
\langle\widetilde{\Omega}_{\pm}\mid e^{-\beta \widetilde{H}_{\mbox{\tiny{QR}}}}\widetilde{\Omega}_{\pm}\rangle
=\langle\widetilde{\Omega}_{\pm}\mid e^{-\beta \widetilde{H}_{\pm}}\widetilde{\Omega}_{\pm}\rangle. 
\end{align*}
We define the constants $E_{\pm}$ by
$$
E_{\pm}=-\lim_{\beta\to\infty}\frac{1}{\beta}
\ln\langle\widetilde{\Omega}_{\pm}\mid e^{-\beta \widetilde{H}_{\pm}}\widetilde{\Omega}_{\pm}\rangle.
$$
Following the arguments in Lemma 2.3 in Ref.\cite{hir99}, as adapted in Ref.\cite{hir25},
we obtain the expression
\begin{equation}
E_{\pm} =
\frac{\hbar\omega_{\mathrm{c}}}{2} 
-\, \frac{\hbar\mathrm{g}^{2}}{\omega_{\mathrm{c}}}
\mp\, \frac{\hbar\omega_{\mathrm{a}}}{2}
\mathrm{exp}
\left[
- 2\frac{\mathrm{g}^{2}}{\omega_{\mathrm{c}}^{2}}
\left( G(\mathrm{g})+1\right) 
\right],
\label{eq:e_pm}
\end{equation}
where the quantity $G(\mathrm{g})$ depends on the coupling strength $\mathrm{g}$
and satisfies $-1\le G(\mathrm{g})\le 0$. 
We consider a situation where neither the ground-state energy nor the first excited energy
has any energy-level crossing for $0<\mathrm{g}<\infty$.
It is noteworthy that there is no energy-level crossing between the ground
and the first excited states \cite{hh14}.
As shown in Figs.\ref{fig:qr_energy_0}-\ref{fig:renqr_energy},
we have such a situation provided that $0<\omega_{\mathrm{a}}\le\omega_{\mathrm{c}}$
and $0<\frac{\mathrm{g}}{\omega_{\mathrm{a}}}<\infty$. 
At least mathematically, this situation occurs for sufficiently large $\frac{\mathrm{g}}{\omega_{\mathrm{a}}}$
\cite{hir15,hir24}. 
Due to the decomposition given by Eq.(\ref{eq:decomposition_0}),
the energies $E_{+}$ and $E_{-}$ are respectively the ground-state energy $E_{0}$
and the first-excited-state energy $E_{1}$ of the transformed quantum Rabi Hamiltonian
$\widetilde{H}_{\mbox{\tiny{QR}}}$ (and thus the quantum Rabi Hamiltonian $H_{\mbox{\tiny{QR}}}$)
since $E_{+}\le E_{-}$: $E_{0}=E_{+}$ and $E_{1}=E_{-}$.

The expressions given in Eq.(\ref{eq:e_pm}) are consistent with Polyakov's statement.
As shown in Ref.\cite{hir25}, we obtain
\begin{equation}
S_{\mathrm{euc}}(q_{\mathrm{cl}})
=
2\hbar\frac{\mathrm{g}^{2}}{\omega_{\mathrm{c}}^{2}}
\left( G(\mathrm{g})+1\right) 
\label{eq:euclidean_action_1}
\end{equation}
for the classical Euclidean action $S_{\mathrm{euc}}(q_{\mathrm{cl}})$
of an effective instanton-like particle. 
For the renormalized quantum Rabi Hamiltonian $H_{\mbox{\tiny{QR}}}^{\mathrm{ren}}$,
Eqs.(\ref{eq:e_pm}) and (\ref{eq:euclidean_action_1}) yield the expressions for
the ground-state energy $E_{0}^{\mathrm{ren}}$ and the first excited-state energy
$E_{1}^{\mathrm{ren}}$ as 
\begin{equation}
  E_{0}^{\mathrm{ren}}
  =
  \frac{\hbar\omega_{\mathrm{c}}}{2}
  -\, \frac{\hbar\omega_{\mathrm{a}}}{2}
  \mathrm{exp}
  \left[
    -\, \frac{S_{\mathrm{euc}}(q_{\mathrm{cl}})}{\hbar}
    \right] 
\label{eq:gse-expression}
\end{equation}
and
\begin{equation}
  E_{1}^{\mathrm{ren}}
  =
  \frac{\hbar\omega_{\mathrm{c}}}{2}
  + \frac{\hbar\omega_{\mathrm{a}}}{2}
  \mathrm{exp}
  \left[
    -\, \frac{S_{\mathrm{euc}}(q_{\mathrm{cl}})}{\hbar}
    \right].  
\label{eq:1se-expression}
\end{equation}
The expressions in Eqs.(\ref{eq:gse-expression}) and (\ref{eq:1se-expression})
are similar to Eq. (12.68) of Ref.\cite{razavy03}.
Based on the results in Refs.\cite{hir99, hir25},
$G(\mathrm{g})\ne -1$ holds for every finite $\mathrm{g}$;
consequently, we obtain the limit
$$
\lim_{\mbox{\tiny{SB}}}
\frac{\hbar\omega_{\mathrm{a}}}{2}
\mathrm{exp}
\left[
  -\, \frac{S_{\mathrm{euc}}(q_{\mathrm{cl}})}{\hbar} 
  \right]
=
\lim_{\mbox{\tiny{SB}}}
\frac{\hbar\omega_{\mathrm{a}}}{2}
\mathrm{exp}
\left[
- 2\frac{\mathrm{g}^{2}}{\omega_{\mathrm{c}}^{2}}
\left( G(\mathrm{g})+1\right) 
\right]
=0 
$$
(see Figs.\ref{fig:qr_energy_2}, \ref{fig:renqr_energy}, and \ref{fig:renqr_energy_r}).  
We find that the energy levels $E_{0}^{\mathrm{ren}}$ and $E_{1}^{\mathrm{ren}}$
form a degenerate ground state at $\frac{\mathrm{g}\,\,}{\omega_{\mathrm{a}}}=\infty$: 
\begin{align}
E_{0}^{\mathrm{ren}}=\frac{\hbar\omega_{\mathrm{c}}}{2}=E_{1}^{\mathrm{ren}}.
\label{eq:degeneracy}
\end{align}
On the other hand, once the ratio $\frac{\mathrm{g}\,\,}{\omega_{\mathrm{a}}}$
becomes finite (i.e., $\frac{\mathrm{g}\,\,}{\omega_{\mathrm{a}}}<\infty$),
Polyakov's bifurcation takes place:
$$
E_{0}^{\mathrm{ren}}=
\frac{\hbar\omega_{\mathrm{c}}}{2} 
-\,
\frac{\hbar\omega_{\mathrm{a}}}{2}
  \mathrm{exp}
  \left[
    -\, \frac{S_{\mathrm{euc}}(q_{\mathrm{cl}})}{\hbar}
    \right]
<\frac{\hbar\omega_{\mathrm{c}}}{2} 
+ \frac{\hbar\omega_{\mathrm{a}}}{2}
  \mathrm{exp}
  \left[
    -\, \frac{S_{\mathrm{euc}}(q_{\mathrm{cl}})}{\hbar}
    \right]
=E_{1}^{\mathrm{ren}}.
$$
In particular the inequality becomes
$E_{0}^{\mathrm{ren}}=
\frac{\hbar\omega_{\mathrm{c}}}{2} 
-\,
\frac{\hbar\omega_{\mathrm{a}}}{2}<
\frac{\hbar\omega_{\mathrm{c}}}{2} 
+ \frac{\hbar\omega_{\mathrm{a}}}{2}
=E_{1}^{\mathrm{ren}}$ at $\frac{\mathrm{g}\,\,}{\omega_{\mathrm{a}}}=0$. 
Therefore, we can conclude that the tunneling term
$\frac{\hbar\omega_{\mathrm{a}}}{2}\mathrm{exp}\left[-\, \frac{S_{\mathrm{euc}}(q_{\mathrm{cl}})}{\hbar}\right]$
induced by the effective instanton-like particle is consistent with Polyakov's findings
(see Figs.\ref{fig:qr_energy_2}, \ref{fig:renqr_energy}, and \ref{fig:renqr_energy_r}).

In experimental observations, energy levels are typically determined by measuring energy differences
(e.g., Fig. 2b in Ref.\cite{cai22}), such as through photon emission.
It is therefore meaningful to express the classical Euclidean action $S_{\mathrm{euc}}(q_{\mathrm{cl}})$
in terms of the energy gap between the ground and first excited states.
In the same way that Eq. (12.69) is derived in Ref.\cite{razavy03},
Eqs.(\ref{eq:gse-expression}) and (\ref{eq:1se-expression})
lead to the expression
\begin{align}
S_{\mathrm{euc}}(q_{\mathrm{cl}})
=-\hbar\ln\left[\frac{E_{1}^{\sharp}-E_{0}^{\sharp}}{\hbar\omega_{\mathrm{a}}}\right],
\label{eq:euclidean_action_2}
\end{align}
where $E_{i}^{\sharp}$ is $E_{i}$ or $E_{i}^{\mathrm{ren}}$ for $i=0, 1$. 

We, for example, set the double-well potential as $V(x)=C_{\mathrm{dw}}(x^{2}-q_{0}^{2})^{2}$ 
with a positive constant $C_{\mathrm{dw}}$.
The classical Euclidean action for the macrosystem Hamiltonian $H_{\mbox{\tiny{MS}}}$,
defined by Eq.(\ref{eq:ms-Hamiltonian}),
is then given by Eq.(2.22) of Ref.\cite{col85} or Eq. (3.36) of Ref.\cite{as10}; hence, we obtain the approximation
$$
S_{\mathrm{euc}}(q_{\mathrm{cl}})
\approx \int_{-q_{0}}^{+q_{0}}\sqrt{2V(x)}\, dx. 
$$
This approximation and Eq.(\ref{eq:euclidean_action_2}) yield 
\begin{align}
q_{0}\approx
\left\{
-\,\frac{3\hbar}{4\sqrt{2C_{\mathrm{dw}}}}\ln\left[
\frac{E_{1}^{\sharp}-E_{0}^{\sharp}}{\hbar\omega_{\mathrm{a}}}
  \right]\right\}^{1/3}, 
\label{eq:q_0}
\end{align}
which reveals that the coordinates of the minima, $\pm q_{0}$, 
diverge from each other and tend to $\pm\infty$, respectively, in (LMT1),
although further analysis of the behavior of the logarithmic term 
is necessary in (LMT2).

\section{Conclusion and Discussion}

We have demonstrated that the transformed quantum Rabi model
effectively describes macroscopic quantum tunneling (MQT).
Our mathematical analysis revealed that the bifurcation proposed by Polyakov
arises for recovery from spontaneous symmetry breaking.
This suggests that Polyakov's bifurcation is potentially observable
in quantum simulations,
which raises further questions regarding the realization and
detectability of the effective instanton-like particle in such systems.
Furthermore, we have shown that $\mathcal{N}=2$ supersymmetry (SUSY)
emerges as a special case of the reverse process of
Polyakov's bifurcation.

Let us point out one specific problem regarding these questions. 
Since the term
$\frac{\hbar\omega_{\mathrm{a}}}{2}
  \mathrm{exp}
  \left[
    -\, \frac{S_{\mathrm{euc}}(q_{\mathrm{cl}})}{\hbar}
    \right]$
  in Eqs.(\ref{eq:gse-expression}) and (\ref{eq:1se-expression}) vanishes
  in both (LMT1) and (LMT2), which implies Eq.(\ref{eq:degeneracy}),
  the tunneling of single-mode bosons disappears
  when spontaneous $\mathbb{Z}_{2}$-symmetry breaking occurs.
  Regarding (LMT1), this vanishing mathematically occurs after taking the limit;
  however, as shown in Figs.\ref{fig:renqr_energy_1} and \ref{fig:renqr_energy},
  the decay is rapid.
  Of interest is the behavior of the classical Euclidean action $S_{\mathrm{euc}}(q_{\mathrm{cl}})$, 
  namely, the existence of instanton-like configurations, in these two limits. 
  Furthermore, if such configurations exist, how they cease to facilitate tunneling must be investigated,
  as demonstrated, for instance, by Eq.(\ref{eq:q_0}) in (LMT1).
  Meanwhile, the spontaneous $\mathbb{Z}_{2}$-symmetry breaking should produce
  the Nambu-Goldstone fermion \cite{volkov73,salam74,wit81,wit82}.
  It is also of interest whether a relation exists between the instanton-like configurations
  in (LMT1) or (LMT2) and the Nambu-Goldstone fermion described in Refs.\cite{hir24,hir25}
  --- in particular, whether there is a general mechanism
  by which tunneling penetration chooses SUSY.

We expect that simulations based on the quantum Rabi model will provide
a robust platform for investigating phenomena caused by instantons
both theoretically and experimentally.
These findings open up possibilities for extending
the current framework to spin-boson models,
potentially paving the way for further considerations in quantum field theory.

\begin{acknowledgments}
This work was supported by the Quantum and Spacetime Research Institute (QuaSR), Kyushu University.  
The author thanks Shiro Saito and Yutaka Tabuchi for the fruitful discussions.
This Perspective is dedicated to Pavel Exner, Herbert Spohn, and Valentin Zagrebnov
on the occasion of their 80th birthdays.
\end{acknowledgments}

\section*{DATA AVAILABILITY}
All energy spectra data in the figures can be generated using Python code with QuTiP \cite{nori1, nori2}.

\bibliography{hirokawa}

\end{document}